\documentstyle[11pt,aaspp4]{article}

\input{psfig.sty}

\begin{document}

\title{Changes in the X-ray Emission from the Magnetar Candidate 1E~2259$+$586
during its 2002 Outburst}

\author{
P.M.~Woods\altaffilmark{1,2},
V.M.~Kaspi\altaffilmark{3,4}, 
C.~Thompson\altaffilmark{5},
F.P.~Gavriil\altaffilmark{3},
H.L.~Marshall\altaffilmark{4},
D.~Chakrabarty\altaffilmark{4},
K.~Flanagan\altaffilmark{4},
J.~Heyl\altaffilmark{6,7}, and
L.~Hernquist\altaffilmark{6}
}

\altaffiltext{1}{Universities Space Research Association; 
peter.woods@nsstc.nasa.gov}
\altaffiltext{2}{National Space Science and Technology Center, 320 Sparkman Dr. 
Huntsville, AL 35805}
\altaffiltext{3}{Department of Physics, Rutherford Physics Building, McGill 
University, 3600 University St., Montreal, Quebec, H3A~2T8, Canada}
\altaffiltext{4}{Center for Space Research,
Massachusetts Institute of Technology, Cambridge, MA 02139}
\altaffiltext{5}{Canadian Institute for Theoretical Astrophysics, 60 St.\
George Street, Toronto, ON M5S 3H8, Canada}
\altaffiltext{6}{Harvard-Smithsonian Center for Astrophysics, 60 Garden St.,
Cambridge, MA 02138}
\altaffiltext{7}{Chandra Fellow, Current Address: Department of Physics and
Astronomy, 6224 Agricultural Road, Vancouver BC V6T 1Z1, Canada}

\begin{abstract}

An outburst of more than 80 individual bursts, similar to those seen from Soft
Gamma Repeaters (SGRs), was detected from the Anomalous X-ray Pulsar (AXP)
1E~2259$+$586 in 2002 June.  Coincident with this burst activity were gross
changes in the pulsed flux, persistent flux, energy spectrum, pulse profile and
spin down of the underlying X-ray source.  We present {\it Rossi X-ray Timing
Explorer} and {\it X-ray Multi-Mirror Mission} observations of 1E~2259$+$586
that show the evolution of the aforementioned source parameters during and
following this episode and identify recovery time scales for each. 
Specifically, we observe an X-ray flux increase (pulsed and phase-averaged) by
more than an order of magnitude having two distinct components.  The first
component is linked to the burst activity and decays within $\sim$2 days during
which the energy spectrum is considerably harder than during the quiescent
state of the source.  The second component decays over the year following the
glitch according to a power law in time with an exponent $-0.22 \pm 0.01$.  The
pulsed fraction decreased initially to $\sim$15\% RMS, but recovered rapidly to
the pre-outburst level of $\sim$23\% within the first three days.  The pulse
profile changed significantly during the outburst, and recovered almost fully
within two months of the outburst.  A glitch of size $\Delta\nu_{\rm max}/\nu =
(4.24 \pm  0.11) \times 10^{-6}$ was observed in 1E~2259$+$586 that preceded
the observed burst activity.  The glitch could not be well fit with a simple
partial exponential recovery.  An exponential rise of $\sim$20\% of the
frequency jump with a time scale of $\sim$14 days results in a significantly
better fit to the data, however contamination from a systematic drift in the
phase of the pulse profile cannot be excluded.  A fraction of the glitch
($\sim$19\%) recovered in a quasi-exponential manner having a recovery time
scale of $\sim$16 days.  The long-term post-glitch spin-down rate decreased in
magnitude relative to the pre-glitch value.  The changes in the source
properties of 1E~2259$+$586 during its 2002 outburst are shown to be
qualitatively similar to changes seen during/following burst activity in two
SGRs, thus further solidifying the common nature of SGRs and AXPs as
magnetars.  The changes in persistent emission properties of 1E~2259$+$586
suggest that the star underwent a plastic deformation of the crust that
simultaneously impacted the superfluid interior (crustal and possibly core
superfluid) and the magnetosphere.  Finally, the changes in persistent emission
properties coincident with burst activity in 1E~2259$+$586 enabled us to infer
previous burst active episodes from this and other AXPs.  The non-detection of
these outbursts by all-sky gamma-ray instruments suggests that the number of
active magnetar candidates in our Galaxy is larger than previously thought. 

\end{abstract}


\newpage

\section{Introduction}

Anomalous X-ray Pulsars (AXPs) and Soft Gamma Repeaters (SGRs) are two
intriguing classes of isolated neutron stars, very likely magnetars, whose
bright X-ray emission is powered by the decay of their strong magnetic fields. 
When the common nature of AXPs and SGRs was first proposed by Thompson \&
Duncan (1996) with both being magnetars, the observational evidence linking
them was tenuous.  A major advance in connecting these two classes came when
slow pulsations and rapid spin down, defining characteristics of AXPs, were
discovered from SGR~1806$-$20 (Kouveliotou et al.\ 1998).  Since then, further
observational similarities among AXPs and SGRs have been established.  For
example, Marsden \& White (2001) performed a systematic analysis of SGR and AXP
spectral data that showed the two classes form a continuum in spectral hardness
versus spin-down rate where the SGRs have harder spectra and faster spin-down
rates than the spectrally softer, slower braking AXPs.  Similarly, the timing
noise strength in SGRs and AXPs appears to be correlated with spin-down rate
(Woods et al.\ 2000; Gavriil \& Kaspi 2002).  For a review of SGRs and AXPs,
see Kouveliotou (2003) and Mereghetti et al.\ (2002), respectively.  Until
recently, the most prominent characteristic distinguishing the two groups was
the emission of soft, bright (up to 10$^{44}$ ergs s$^{-1}$) bursts of soft
gamma-rays from the SGRs and {\it not} the AXPs.  In fact, it was this
extraordinary property of SGRs which led to doubt within the community that
AXPs and SGRs were of the same nature.  This uncertainty was removed when
SGR-like bursts were recently discovered from at least one AXP (Kaspi et al.\
2003), and probably one other (Gavriil, Kaspi \& Woods 2002).

SGR-like bursts from the direction of an AXP were first detected from the
source 1E~1048.1$-$5937 (Gavriil, Kaspi \& Woods 2002).  A single weak SGR-like
burst was detected during each of two {\it Rossi X-ray Timing Explorer (RXTE)}
Proportional Counter Array (PCA) monitoring observations of this AXP separated
by two weeks.  However, the identification of 1E~1048.1$-$5937 as the burst
source could not be made unambiguously due to the lack of imaging capability
with the PCA.  Interestingly, the quiescent properties of this AXP (e.g.\
energy spectrum, pulse profile, timing noise) most closely resemble those of
the SGRs (Kaspi et al.\ 2001), making this AXP the most SGR-like of its class.

The second detection of SGR-like bursts was recorded from the AXP 1E~2259$+$586
on 2002 June 18.  This is the least SGR-like of the AXPs in terms of its
quiescent source properties.  Unlike the two weak bursts observed earlier from
1E~1048.1$-$5937, this AXP showed a major SGR-like outburst, or collection of
bursts (Kaspi et al.\ 2003).  In total, more than 80 bursts were detected
within 3 hours of observing time.  A detailed study of these bursts will be
presented in a companion paper (Gavriil et al.\ 2003).  In addition to emitting
these hard X-ray bursts, several parameters of the persistent source changed in
conjunction with this outburst, thereby confirming the AXP as the source of the
burst emission (Kaspi et al.\ 2003).

Here, we present {\it X-ray Multi-Mirror Mission (XMM-Newton)} and {\it RXTE}
PCA observations of the persistent X-ray flux from 1E~2259$+$586 before, during
and after the 2002 June outburst.  We quantify the changes of the spectral and
temporal properties of the X-ray source as well as the time scales for their
recovery.  We compare the changes observed in this AXP to dynamic behavior seen
in the persistent emission of SGR~1900$+$14 (Woods et al.\ 2001) and
SGR~1627$-$41 (Kouveliotou et al.\ 2003) during episodes of intense burst
activity.  Finally, we present a possible explanation for the observed behavior
in 1E~2259$+$586 within the context of the magnetar model.

\section{{\bf {\it XMM-Newton}} Observations}

The results presented in this section were obtained from observations of
1E~2259$+$586 with the telescopes aboard the {\it XMM-Newton} observatory
(Jansen et al.\ 2001).  This observatory is comprised of three co-aligned X-ray
telescopes.  The focal plane instruments are one EPIC PN camera (Str\"uder et
al.\ 2001) and two EPIC MOS cameras (Turner et al.\ 2001).  All instruments are
sensitive to X-rays between 0.2 and 15.0 keV.  The PN camera has an effective
area of $\sim$1400 cm$^{2}$ at 1.5 keV while the MOS cameras each have areas
of $\sim$500 cm$^{2}$ at 1.5 keV.  The focused X-ray beam for the telescopes
serving each MOS camera is split between the focal plane instrument and the
Reflection Grating Spectrometer (RGS [den Herder et al.\ 2001]).

The {\it XMM-Newton} observatory observed 1E~2259$+$586 five times during
2002.  Three pointings were focused on different portions of CTB~109, the SNR
surrounding 1E~2259$+$586.  The primary scientific objective of the remaining
two observations was the central point source.  In each of the three CTB~109
pointings, the AXP is within the field-of-view at large off-axis angles
(10$-$13$^{\prime}$).  Results from all five observations on the central source
will be presented here.  Analysis of the data from the remnant will be
presented elsewhere (Sasaki et al., in preparation).

The first {\it XMM-Newton} observation of 1E~2259$+$586 was one of the remnant
pointings, carried out on 2002 January 22.  The second was centered on the AXP
and was performed on 2002 June 11, fortuitously one week prior to the outburst
of 1E~2259$+$586.  Following the outburst, a Target of Opportunity (ToO) was
declared and the source was reobserved three days later on 2002 June 21.  The
remaining two observations of CTB~109 were carried out on 2002 July 09.  See
Table 1 for further details on these observations.  Hereafter, each {\it
XMM-Newton} observation will be referred to by the identifier label assigned to
it in the first column of Table 1.


\begin{table}[!h]
\begin{minipage}{1.0\textwidth}
\begin{center}
\caption{{\it XMM-Newton} observation log for 1E~2259$+$586.} 
\vspace{10pt}
\begin{tabular}{ccccccc} \hline \hline

 Name & {\it XMM-Newton} & Relation to    &  Date$^{a}$ & PN Exposure$^{b}$ & PN Data Mode$^{c}$ & AXP Offset$^{d}$   \\
      & Obsid            & Burst Activity &  (MJD TDB)  & (ksec)            &                    &  (arcmin)            \\\hline
 
 Obs1 & 0057540101  & Before  &  52296.791  &  10.7  & Extended &   11.2    \\
 Obs2 & 0038140101  & Before  &  52436.546  &  24.9  & Small    &    1.7    \\
 Obs3 & 0155350301  & After   &  52446.449  &  18.5  & Small    &    1.7    \\
 Obs4 & 0057540201  & After   &  52464.369  &  12.3  & Extended &   13.0    \\
 Obs5 & 0057540301  & After   &  52464.606  &  12.4  & Extended &    9.7    \\

\hline\hline
\end{tabular}
\end{center}
\noindent$^{a}$ Start time of each observation.  Note that the outburst began 
on date 52443.65 MJD. \\
\noindent$^{b}$ Exposure times quoted reflect on-source times after filtering 
of flares, etc. for spectral and temporal analysis. \\
\noindent$^{c}$ Extended = Extended Full Frame Mode; Small = Small Window Mode. \\
\noindent$^{d}$ The angular distance of 1E~2259$+$586 from the optical axis of
the EPIC mirrors.
\end{minipage}\hfill
\end{table}

The two MOS cameras were operated in full-frame mode for all but one
observation (Obs3 $-$ Small Window for MOS1) in order to study the SNR.  The
frame time for the full-frame data mode is 2.6 s which causes severe pile-up
for the AXP ($\gtrsim$50\%). Therefore, these data are not considered further
here.

The PN camera was operated in small window mode for both Obs2 and Obs3, the two
observations closely bracketing the 2002 June outburst in time.  The frame time
for the PN camera in this mode is $\simeq$5.86 ms, allowing detailed study of
the pulsed emission and a search for burst emission.  For the on-axis count
rate of the AXP, the dead-time fraction is 30\%, but the pile-up fraction is
only 0.03\%.  The three off-axis pointings (Obs1, Obs4, and Obs5) were carried
out in Extended Full-Frame mode which has a time resolution of 200 ms.  The
high off-axis angles for the AXP reduced the count rate by a factor 2$-$3.  The
different data mode and reduction in count rate lessened the dead-time to 2\%,
but the pile-up fraction increased to 3\%.

The RGS data from Obs2 and Obs3 were acquired in spectroscopy mode with 5.7 s
time resolution and excellent energy resolution ($\Delta E/E$ = 10$-$20).  The
coarse time resolution allowed for only phase-averaged spectral analysis (see
\S2.3).

All data were processed using the {\it XMM-Newton} Science Analysis 
System\footnote{http://xmm.vilspa.esa.es/external/xmm\_sw\_cal/sas.shtml} (SAS)
v5.4.1.  For the PN data, the script {\tt epchain} was run on the Observation
Data Files.  This script processes the data for use in higher-end analysis
tools.  A light curve of the full field-of-view (FOV) excluding the bright
central source was constructed (0.5$-$10.0 keV) and inspected for times of high
background.  We chose a threshold of 2 times the nominal background to define
regions of high background.  Periods of high background constituted 0$-$33\% of
each data set and were eliminated from further analysis.

\subsection{Burst Search}

We have used the data from the PN camera to search for burst emission from
1E~2259$+$586 during the {\it XMM-Newton} observations.  Using the SAS tool
{\tt evselect}, events from within 10$^{\prime \prime}$ and 12.5$^{\prime
\prime}$ radii circles around the position of the AXP were extracted for the
on-axis and off-axis (remnant) pointings, respectively.  Following the standard
filtering procedures for {\it XMM-Newton} PN data, grades $\ge$12, and flag
values equal 0 were retained.  Next, we filtered the event list on energy
between 0.5 and 12.0 keV to optimize the signal-to-noise ratio for burst
emission.

Light curves were constructed for each of the {\it XMM-Newton} observations at
the nominal time resolution (6 ms for Obs2 and Obs3 and 200 ms for the others),
0.1 s resolution (Obs2 and Obs3), 1 s, and 10 s resolution.  No significant
burst emission was detected on these time scales during any of these
observations.  This is consistent with the absence of bursts in the more
densely sampled {\it RXTE} data (Gavriil et al.\ 2003) that bracket these {\it
XMM-Newton} observations.  Bursts were recorded from 1E~2259$+$586 only on 2002
June 18.

\subsection{Pulse Timing Analysis}

Using the same procedures as those described above for the burst search, source
event lists from the PN data were extracted to study the pulsed emission from
1E~2259$+$586.  In order to optimize the signal-to-noise ratio for the pulsed
emission, the event list was filtered on energy between 0.5 and 7.0 keV.

There is a known problem with the time tags of {\it XMM-Newton} PN data in
which there can be sudden jumps in the photon arrival times of integer second
values (W.~Becker, private communication).  It is still not understood at what
stage of the processing these ``time jumps'' distort the time tags, only that
they do occur.  We searched for and identified four time jumps within two of
the five data sets (Obs 2 and 3).  All time jumps were identified in the two
pointings where the PN camera was operated in small window mode.  To find these
time jumps we calculated the modulus of each event time stamp with the mean
frame time of the corresponding data set.  The frame time varied with data mode
(small window vs. extended full frame) as well as between and within
observations having identical data modes.  The latter effect is believed to be
due to temperature variations in the on-board electronics ({\it XMM-Newton}
Helpdesk, private communication).

The time jumps in each data set were obvious after plotting the frame time
residuals versus the time of each event.  Time jumps manifested themselves as
discontinuities in the frame time residuals plot.  These discontinuities were
identified and corrected for by adding or subtracting an integer number of
seconds to the data following the jump until the frame time residuals matched
precisely across the boundary.  The time corrections we applied to the data
were verified by the pulsar data.  In the uncorrected pulsar data set, phase
jumps were detected that were consistent with being equal to the time
corrections required by the frame time inconsistencies.  Using the time
corrected data set, the phase jumps in the pulsar data disappeared.  Therefore,
we are secure about the relative timing of the corrected {\it XMM-Newton} PN
data set.  However, the absolute timing of these events will require a better
understanding of the origin of the time jump problem. 

The corrected time tags were next converted to the Solar system barycenter
using the SAS tool {\tt barycorr}.  Assuming that the remaining uncertainty in
the corrected time tags is a small integer number of seconds, the propagated
error in the barycenter correction applied to the data is insignificant with
respect to the precision with which we can measure the pulse frequency.

For the two point-source observations in small window mode (Obs2 and Obs3), the
data were binned at twice the nominal frame time of each observation and a Fast
Fourier Transform was constructed of the light curves (0.5$-$7.0 keV).  The
only significant power detected between 4 $\times$ 10$^{-4}$ Hz and the Nyquist
frequency (44 Hz) is from the first 7$-$8 harmonics of the pulsar.  The
3$\sigma$ upper limit to the mean fractional power in the frequency range
0.001$-$1.0 Hz (minus the pulsar frequency and harmonics) is 3 $\times 10^{-3}$
and 2 $\times 10^{-3}$ (RMS/MEAN)$^2$ Hz$^{-1}$, for Obs2 and Obs3
respectively.  These limits are orders of magnitude lower than typical
broadband noise power levels seen in accreting X-ray pulsars, even so-called
``quiet'' accreting pulsars such as 4U~1626$-$67 (Owens, Oosterbroek, \& Parmar
1997; Shinoda et al.\ 1990).

The precise pulse frequencies for each pointing were determined using a phase
folding technique.  For each data set, a template pulse profile was constructed
by folding the data on the frequency as determined from the peak power of the
fundamental frequency in the power density spectrum.  Next, the data set was
split into $\approx$10$^{3}$ s segments and each segment was folded with the
same frequency.  The pulse profiles from each segment were cross-correlated
with the template pulse profile and a phase offset was measured.  The resulting
phase differences for each segment were fit to a line and the slope was used to
correct the frequency. Using the refined frequency, a new template pulse
profile was constructed and the procedure for refining the pulse frequency was
repeated.  The epochs and frequencies measured for each {\it XMM-Newton}
observation are listed in Table 2.  Since two of the CTB~109 observations were
performed sequentially, we determined a single frequency for the combined
observation.  The spin frequencies measured in the {\it XMM-Newton}
observations are consistent with the much more precise spin ephemeris measured
using the {\it RXTE} data (see \S3.1).


\begin{table}[!h]
\begin{minipage}{1.0\textwidth}
\begin{center}
\caption{Pulse frequencies and pulsed fractions of 1E~2259$+$586 measured 
using {\it XMM-Newton} PN data.} 
\vspace{10pt}
\begin{scriptsize}
\begin{tabular}{cccccccc} \hline \hline

 Observation &  Epoch                & Frequency$^{a}$ & & & Pulsed Fraction        \\
             & (MJD TDB)  	     &   (Hz)          & 0.3$-$1.0 keV & 1.0$-$2.0 keV & 2.0$-$5.0 keV & 5.0$-$12.0 keV & 2.0$-$10.0 keV    \\\hline
 
 Obs1        & 52296.791$-$52296.932 & 0.14328688(55)  &  0.169(15) &  0.195(6)  &  0.230(10) &  0.314(73)  &  0.232(10)   \\
 Obs2        & 52436.546$-$52436.983 & 0.14328705(13)  &  0.215(6)  &  0.225(3)  &  0.234(4)  &  0.284(23)  &  0.234(4)    \\
 Obs3        & 52446.449$-$52446.755 & 0.14328746(9)   &  0.168(5)  &  0.200(2)  &  0.223(3)  &  0.189(13)  &  0.220(3)    \\
 Obs4$^{b}$  & 52464.369$-$52464.534 & 0.14328771(13)  &  0.166(15) &  0.180(7)  &  0.225(11) &  0.294(50)  &  0.230(10)   \\
 Obs5$^{b}$  & 52464.606$-$52464.761 & 0.14328771(13)  &  0.139(13) &  0.168(6)  &  0.225(8)  &  0.339(38)  &  0.230(8)    \\

\hline\hline
\end{tabular}
\end{scriptsize}
\end{center}
\noindent$^{a}$ Numbers in parenthesis indicate the 1$\sigma$ uncertainty in the 
least significant digits of the frequency. \\
\noindent$^{b}$ Frequency derived from combined data set of Obs4 and Obs5 due 
to their close proximity to one another in time.
\end{minipage}\hfill
\end{table}

Using the measured frequencies, the background-subtracted PN data were folded
over different energy bands.  The background was estimated by measuring the
count rate for the different energy bands in a circular extraction region close
to the pulsar and within CTB~109, but away from bright knots within the
remnant.  The normalized pulse profiles for different energy bands are shown in
Figure 1.  Clearly, the pulse profile changed significantly directly after the
outburst.  Subsequent X-ray observations revealed a gradual recovery to the
pre-outburst pulse shape (see \S3.4).



\begin{figure}[!htb]
\centerline{
\psfig{file=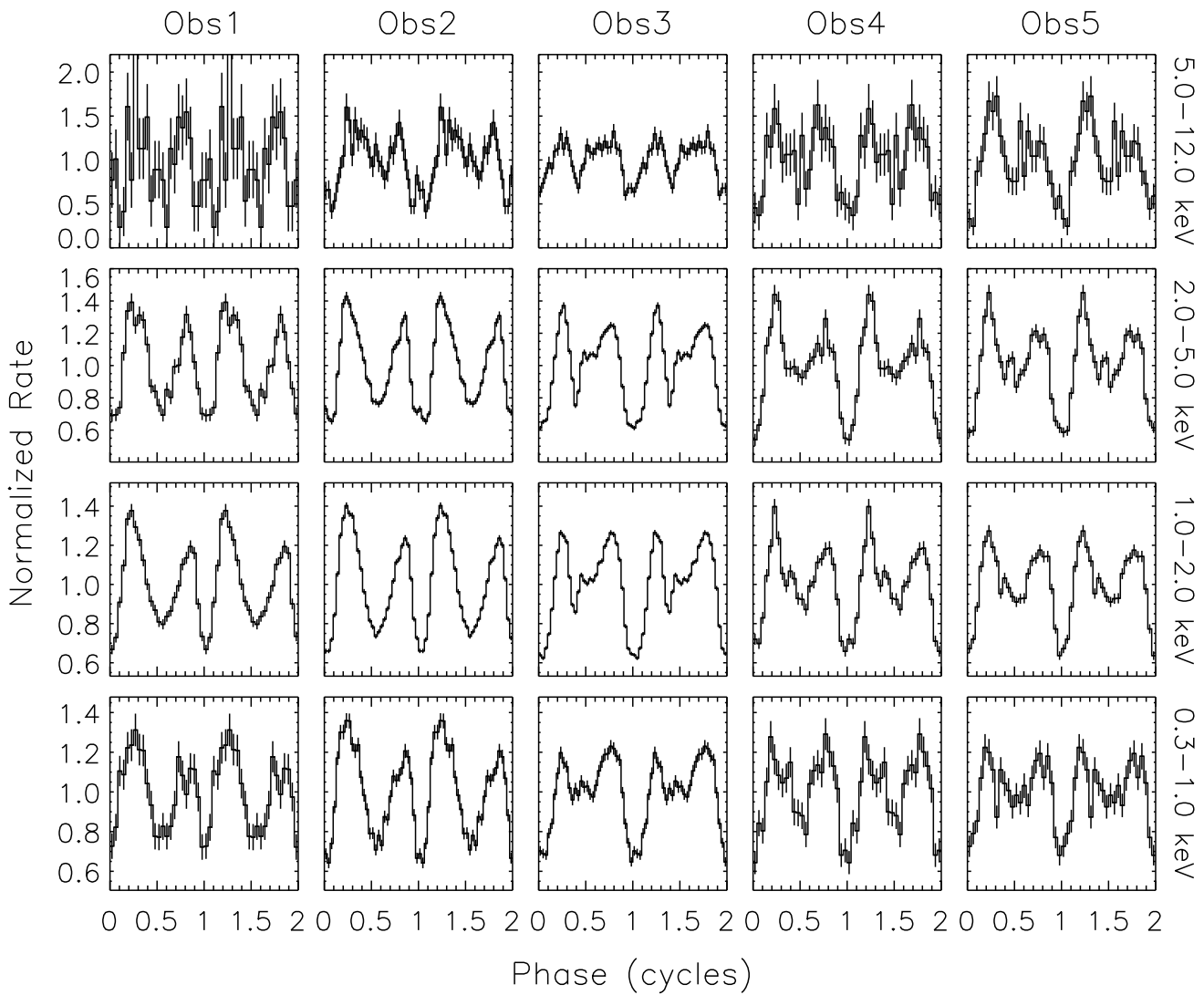,height=6.0in}}
\vspace{-0.05in}

\caption{The folded pulse profile of 1E~2259$+$586 as observed with the {\it
XMM-Newton} PN camera as a function of energy for the all five observations. 
Time increases from left to right.  The burst activity is recorded between Obs2
and Obs3.  See Table 1 for exact times of observations.}

\vspace{11pt}
\end{figure}

The root mean square (RMS) pulsed fractions were calculated for each
observation over several energy bands.  Following van der Klis (1989), the
pulsed fractions were calculated using the first 7 harmonics of the Fourier
representation of the pulse profile according to the equation

\begin{equation}
F_{RMS} = \sqrt{\sum_{k=1}^{7} \frac{\alpha_{k}^{2} + \beta_{k}^{2} - 
   (\sigma_{\alpha_{k}}^{2} + \sigma_{\beta_{k}}^{2})}{2}},
\end{equation}

\noindent where

\begin{eqnarray*}
\alpha_{k} = \frac{1}{N} \sum_{i=1}^{N} r_i \cos{2\pi\phi_i{k}}, &
\displaystyle \beta_{k} = \frac{1}{N} \sum_{i=1}^{N} r_i \sin{2\pi\phi_i{k}}, \\
\sigma_{\alpha_{k}}^{2} = \frac{1}{N^2} \sum_{i=1}^{N} \sigma^{2}_{r_i} 
   \cos^2{2\pi\phi_i{k}},  &
\displaystyle \sigma_{\beta_{k}}^{2} = \frac{1}{N^2} \sum_{i=1}^{N} 
   \sigma^{2}_{r_i} \sin^2{2\pi\phi_i{k}}. \\
\end{eqnarray*}

\noindent Here, $F_{RMS}$ is the pulsed fraction, $k$ refers to the harmonic
number, $i$ refers to the phase bin, $N$ is the total number of phase bins,
$\phi_i$ is the phase, $r_i$ is the count rate in the $i^{\rm th}$ phase bin,
and $\sigma_{x_i}$ is the uncertainty in the count rate of the $i^{\rm th}$
phase bin.  Note that Equation 1 is insensitive to reducing the number of
harmonics used in calculating $F_{RMS}$ for low signal-to-noise pulse profiles
since the statistical noise is subtracted from the total variance.  Seven
harmonics was chosen for $N$ to encompass all statistically significant source
power in the highest signal-to-noise pulse profiles.  The pulsed fractions and
uncertainties are given in Table 2.

Directly following the outburst (Obs3), the pulsed fraction at all energies
dropped relative to the pre-outburst level (Obs2).  The largest change was seen
in the hard band (5.0$-$12.0 keV).  During the {\it XMM-Newton} observations
three weeks after the outburst (Obs4 and Obs5), the broadband pulsed fraction
(2$-$10 keV) had recovered to its pre-outburst value, although the pulse
profile was still significantly different.  The time evolution of the pulsed
fraction is presented in \S3.3 where additional pulsed fraction measurements
made with the {\it RXTE} PCA are reported.

In the observation directly before the outburst (Obs2), the pulsed fraction
does not vary strongly with photon energy.  The energy dependence of the pulsed
fraction is most prominent in the observations following the outburst where the
pulsed fraction increased significantly with energy.  Interestingly, the pulsed
fraction in the soft band (0.3$-$1.0 keV) is higher one week before the
outburst than at any of the other {\it XMM-Newton} epochs before or after.

\subsection{Spectroscopy}

The PN data were used for spectral analysis of 1E~2259$+$586 because of their
excellent signal-to-noise ratio and negligible pile-up.  For the two
observations centered on the point source, events from within a 10$^{\prime
\prime}$ radius circle around the position of the AXP were extracted using the
SAS tool {\tt evselect}.  Following the standard procedures for {\it
XMM-Newton} PN data, these events were filtered for grades $\ge$4, and flag
values equal 0 were retained for the source spectrum.  Similarly, a background
spectrum was extracted using the same filtering criteria from nearby circular
regions of 10$^{\prime \prime}$ radius for each observation identical to the
regions used for the pulsed fraction analysis.  The total number of {\it
source} counts accumulated for each spectrum was 259,000 and 373,000 for the
pre-burst (Obs2) and post-burst (Obs3) observations, respectively.

The three observations of CTB~109 included the AXP, but at large off-axis
angles (10$-$13$^{\prime}$).  At off-axis positions of $\sim$10$^{\prime}$, the
point spread function of the EPIC mirrors is significantly broader than on
axis.  Due to this effect, a larger source extraction radius (12.5$^{\prime
\prime}$) was used.  Background spectra were extracted from circular regions of
radius $\sim$22$^{\prime \prime}$ with centers at off-axis positions similar
to the source extraction region.  The larger extraction radii were chosen to
increase the number of background counts and improve the accuracy of the
background subtraction.  Furthermore, the background spectra were extracted
from the same chip.  Due to the off-axis positions of the AXP and the lower
exposures of the CTB~109 observations, the total number of source counts
extracted in each spectrum was much lower, between 38,000 and 56,000.

Each spectrum was grouped such that there was a minimum of 25 counts per bin. 
The spectra were fit individually using {\bf
XSPEC}\footnote{http://heasarc.gsfc.nasa.gov/docs/xanadu/xspec/} v11.2.0.  Due
to the high column along the line of sight to this source, the observed counts
in Pulse Height Amplitude (PHA) channels corresponding to $E < 0.6$ keV are
dominated by the low PHA tails of events whose true energies are above 1 keV. 
We, therefore, restrict our spectral fits to the data in the range 0.6$-$12.0
keV.  Fits to single component models (blackbody, power law and bremsstrahlung)
modified by interstellar absorption did not yield statistically acceptable
fits.  The resulting reduced $\chi^2$ values to the single component models
were $\gtrsim$4.  We next tried the standard AXP two-component spectral model
of a blackbody plus a power law modified by interstellar absorption.  We
obtained good fits to all five data sets.  The results of these fits are given
in Table 3.  The X-ray spectrum of 1E~2259$+$586 during Obs2 and the residuals
from the best fit are shown in Figure 2.  Formally, the fit to the Obs3
spectrum is not statistically acceptable.  However, the residuals between 0.8
and 3.0 keV constitute the majority of the total $\chi^2$ and are at the few
percent level.  These residuals are very likely due to uncertainties in the
instrumental response (Haberl et al.\ 2003).  Introducing a 3\% systematic
error in the spectral model reduces $\chi^2_{\nu}$ to unity.  The errors quoted
in Table 3 are inflated according the systematic error.  No narrow line
features (absorption or emission) are evident in any of the phase-averaged
spectra.  Within the energy range 0.9$-$2.0 keV, the 90\% confidence limit on a
narrow line feature of the order of the PN response function ($\sim$50 eV FWHM)
is $\sim$10 eV for the equivalent width.  Between 2.0 and 7.0 keV, the response
function increases up to $\sim$130 eV and the line limit increases from
$\sim$10 eV to $\sim$70 eV.



\begin{figure}[!htb]
\centerline{
\psfig{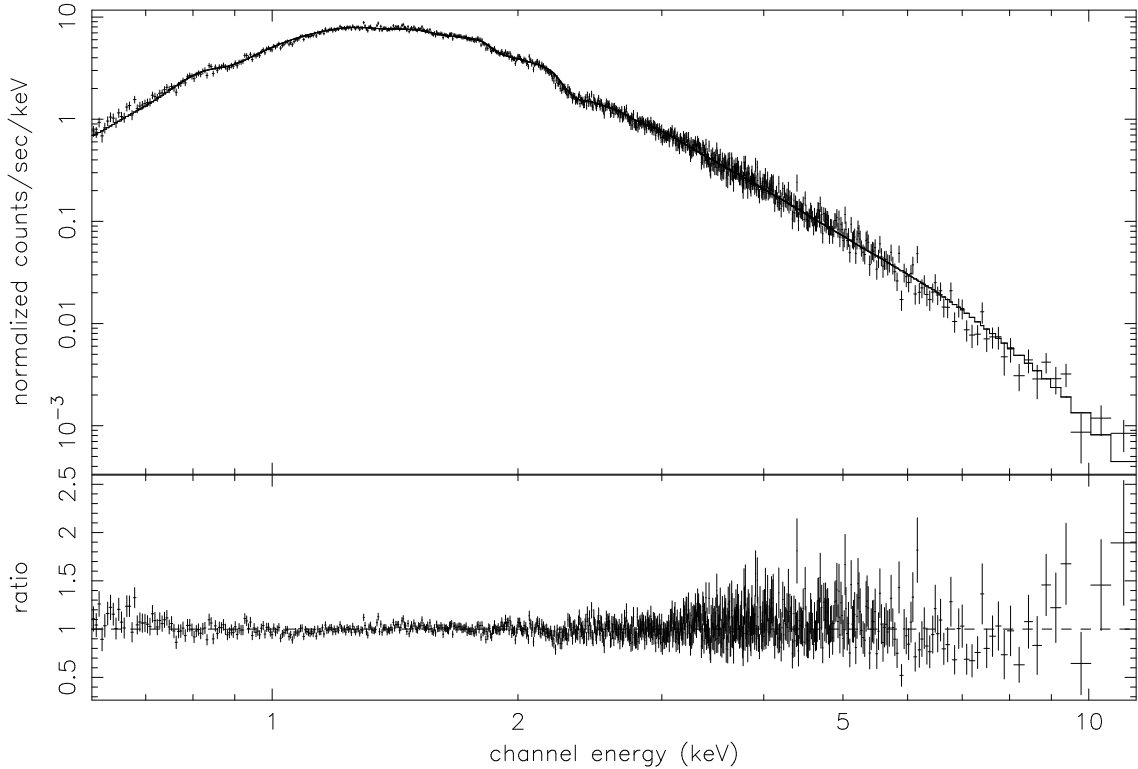}}
\vspace{-0.05in}

\caption{The phase-averaged X-ray spectrum of 1E~2259$+$586 as observed with
the {\it XMM-Newton} PN camera during the observation one week prior to the
2002 June outburst (Obs2).  The ratio of the data to the best-fit model (see
Table 3) folded through the instrumental response is shown in the bottom
panel.}

\vspace{11pt}
\end{figure}


\begin{table}[!h]
\begin{minipage}{1.0\textwidth}
\begin{center}
\caption{Phase-averaged spectral fit parameters of 1E~2259$+$586 from 
{\it XMM-Newton} PN data.} 
\vspace{10pt}
\begin{tabular}{cccccc} \hline \hline

Parameter$^{a}$ 	      &  Obs1     &  Obs2     &  Obs3    &  Obs4     &  Obs5     \\\hline
 
$N_{H}$ (10$^{22}$ cm$^{-2}$) & 1.096(24) & 1.098(12) & 1.035(8) & 0.953(24) & 0.937(20) \\
$kT$ (keV)		      & 0.488(14) & 0.411(4)  & 0.517(5) & 0.537(19) & 0.548(13) \\
$\Gamma$   		      & 4.04(8)   & 4.10(3)   & 3.59(2)  & 3.62(7)   & 3.58(6)   \\
Flux$^{b}$ (10$^{-11}$ ergs cm$^{-2}$ s$^{-1}$)       & 1.24(4) & 1.30(2) & 3.47(3) & 2.01(5) & 2.11(5) \\
Unabs Flux$^{c}$ (10$^{-11}$ ergs cm$^{-2}$ s$^{-1}$) & 1.53(4) & 1.63(2) & 4.17(4) & 2.37(6) & 2.49(5) \\
PL/BB ratio$^{d}$             & 0.70(6)   & 0.43(2)   & 0.74(2)  & 0.93(9)    & 0.79(6)  \\
$\chi^2$/dof                  & 619/578   &  922/851  & 1332/1094 & 606/540  & 599/638 \\

\hline\hline
\end{tabular}
\end{center}
\noindent$^{a}$ Numbers in parentheses indicate the 1$\sigma$ uncertainty in 
the least significant digits of the spectral parameter. Note that these
uncertainties reflect the 1$\sigma$ error for a reduced $\chi^2$ of unity. \\
\noindent$^{b}$ Observed flux from both spectral components 2$-$10 keV. \\
\noindent$^{c}$ Unabsorbed flux from both spectral components 2$-$10 keV. \\
\noindent$^{d}$ The ratio of the power-law flux (2$-$10 keV) to the bolometric
blackbody flux (corrected for absorption).
\end{minipage}\hfill
\end{table}

We next included the RGS data in our spectral analysis.  The RGS spectra for
Obs2 and Obs3 were extracted using standard processing techniques for a point
source.  The data were grouped to 25 counts per spectral bin and ported into
{\bf XSPEC} for simultaneous fitting with the PN data.  Each observation was
fit independently, again to the blackbody plus power law model, and the
measured spectral parameters were consistent with those obtained using the PN
data alone.  The superior energy resolution of the RGS data combined with the
high signal-to-noise data from the PN camera allow us to put even more
constraining limits on the presence of narrow line features.  Within the energy
range 0.8$-$1.75 keV, the 90\% confidence limit on a narrow line feature
($\sim$3$-$10 eV) is $\sim$7 eV for the equivalent width.

We used the RGS data to confirm whether the fit value of $N_H$ was  affected by
the steep power-law component of the spectral model.   Specifically, using just
the PN data, it is difficult to distinguish  between a steep power law that is
strongly absorbed and a flat or  inverted power law (e.g.\ blackbody) that is
observed through modest  absorption.  The Ne-K edge at 0.87 keV is readily
detected in the RGS  data.  Freeing both the $N_H$ and the Ne abundance
relative to solar, $a_{Ne}$, using the {\tt  tbvarabs} model in {\bf XSPEC}, we
find that $a_{Ne}$ = 1.27 $\pm$ 0.24  (at 90\% confidence for one degree of
freedom) and the best-fit value of $N_H$ is very close to the best-fit value
when $a_{Ne}$ is fixed at 1.   Thus, unless the true Ne abundance is
significantly different from the  solar value, the fitted $N_H$ seems accurate
(see Table 3).

Clearly, the energy spectrum of 1E~2259$+$586 hardened following the 2002 June
outburst.  Between the {\it XMM-Newton} observation one week prior to the
outburst and three days following, the photon index became significantly
flatter and the blackbody temperature rose.  However, the blackbody temperature
during Obs1, several months before the outburst, was significantly higher than
the Obs2 temperature and only marginally lower than the post-outburst
temperature, albeit with a considerably lower flux.  This might suggest that
the low temperature measured during Obs2 was an indicator of the impending
outburst.  However, the blackbody temperature measured using the {\it Chandra
X-ray Observatory} in 2000 January was 0.412 $\pm$ 0.006 keV (Patel et al.\
2002), and there was no glitch detected from 1E~2259$+$586 at this time
(Gavriil \& Kaspi 2002).  Moreover, Patel et al.\ (2002) measure a photon index
of 3.6 $\pm$ 0.1, consistent with the post-outburst value.  Other previous
observations of this AXP have shown similar variance in these spectral
parameters (e.g.\ Parmar et al.\ 1998; Marsden \& White 2001), in addition to
the PL/BB flux ratio (Marsden \& White 2001).  Thus, it appears that
1E~2259$+$586 undergoes significant spectral changes independent of large
outbursts and the spectral parameters measured during all of the {\it
XMM-Newton} observations (even those following the outburst) are within the
historical range of these parameters.



\begin{figure}[!htb]
\centerline{
\psfig{file=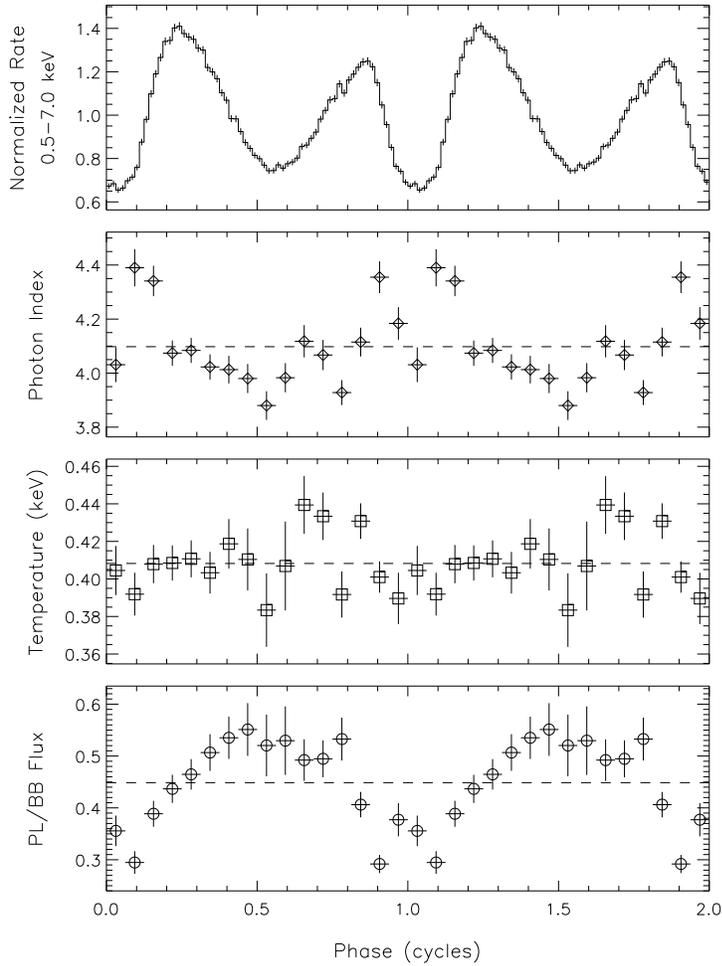,height=5.8in}}
\vspace{-0.05in}

\caption{The phase resolved X-ray spectrum of 1E~2259$+$586 as observed with
the {\it XMM-Newton} PN camera seven days prior to the observed burst activity
(Obs2).  From the top panel down:  the folded pulse profile over the energy
range 0.5$-$7.0 keV, the photon index, the blackbody temperature ($kT$), the
ratio of power-law to blackbody flux.  The horizontal dashed lines in the
bottom three panels denote the average value for each spectral parameter.  For
the ratio of power-law to blackbody flux, the power-law flux was summed over
the energy range 2$-$10 keV and the blackbody was summed over all photon
energies, each corrected for the interstellar absorption (see text for value).}

\vspace{11pt}
\end{figure}



\begin{figure}[!htb]
\centerline{
\psfig{file=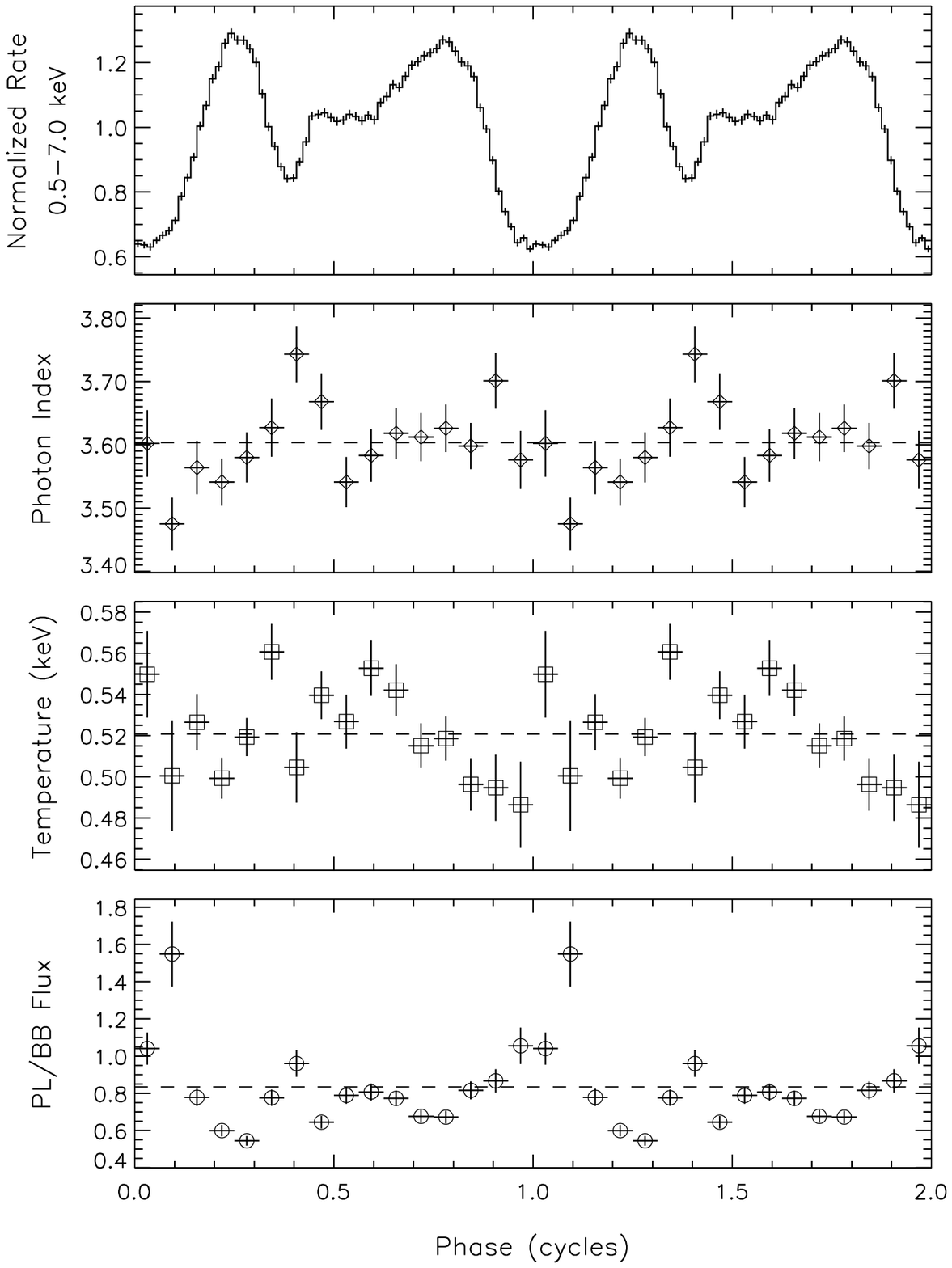,height=5.8in}}
\vspace{-0.05in}

\caption{The phase resolved X-ray spectrum of 1E~2259$+$586 as observed with
the {\it XMM-Newton} PN camera three days after the observed burst activity
(Obs3).  From the top panel down:  the folded pulse profile over the energy
range 0.5$-$7.0 keV, the photon index, the blackbody temperature ($kT$), the
ratio of power-law to blackbody flux.  The horizontal dashed lines in the
bottom three panels denote the average value for each spectral parameter.  For
the ratio of power-law to blackbody flux, the power-law flux was summed over
the energy range 2$-$10 keV and the blackbody was summed over all photon
energies, each corrected for the interstellar absorption (see text for value).}

\vspace{11pt}
\end{figure}

Using the frequencies given in the last section, we extracted spectra for 16
phase intervals per cycle per observation.  For the observation one week prior
to the outburst (Obs2), between 10,000 and 22,000 counts were accumulated per
phase interval.  The spectrum having the least number of total counts (i.e.\
pulse minimum) was grouped such that at least 25 counts were contained in each
data bin.  The grouping for each of the remaining data sets was forced to be
identical to the grouping used for pulse minimum.  As with the phase-averaged
spectrum, we fit a blackbody plus a power law modified by interstellar
absorption.  The sixteen intervals were fit simultaneously with the column
density forced to be the same for all intervals.  The measured column density
is consistent with the phase-averaged value.  All other parameters were allowed
to vary in the fit.  The folded pulse profile, photon index, blackbody
temperature, and ratio of power-law to blackbody flux are shown in Figure 3. 
The average values of the measured spectral parameters are denoted by the
horizontal dashed line shown within each panel.  The flux ratio uncertainties
are determined by propagating the flux errors in each spectral component.  The
probability that the variance in the spectral parameters with phase is purely
statistical is 1.4 $\times$ 10$^{-15}$, 0.10, and $<10^{-16}$ for the photon
index, blackbody temperature, and the flux ratio of the two components.  The
superb statistics of the PN data enabled us to identify both the photon index
and flux ratio as varying significantly as a function of pulse phase whereas
the blackbody temperature is consistent with remaining constant.  

Taken at face value, this observation favors two distinct components to the
1E~2259$+$586 spectrum as has been argued previously (Thompson \& Duncan 1996;
Perna et al.\ 2001).  However, the two components are likely highly correlated
in order to explain the minimal variation of the pulse fraction and shape
versus energy (\"Ozel, Psaltis, \& Kaspi 2001).  Alternatively, a single
thermal component modified by the strong magnetic field (\"Ozel 2001) may be
possible if there is substantial variation in the magnetic field across the
stellar surface.

The same analysis procedure was applied to the data directly following the
outburst.  The folded profile and spectral parameters versus pulse phase are
shown in Figure 4.  Unlike the observation 10 days earlier, the changes in
photon index are only marginally significant (2.9 $\times$ 10$^{-3}$). 
Similarly, the blackbody temperature changes are also marginally significant
(1 $\times$ 10$^{-3}$).  The flux ratio, on the other hand, still shows
significant variability ($<10^{-16}$).  The phase dependence of the flux ratio
data is markedly different than what was seen pre-outburst.

The total number of counts recorded from 1E~2259$+$586 within the off-axis
pointings (Obs1/4/5) is comparable to the counts recorded within only two phase
bins for the on-axis pointings.  Due to the poorer count statistics in these
observations, phase resolved spectroscopy was not performed on these data.

\section{{\bf {\it RXTE}} PCA Observations}

The bursting behavior on 2002 June 18 from 1E~2259$+$586 (Figure 5) was
detected as part of an ongoing monitoring campaign of AXPs spanning the last
six years (e.g.\ Kaspi, Chakrabarty \& Steinberger 1999; Gavriil \& Kaspi
2002).  Follow-up ToO observations of the source with the PCA were immediately
triggered and started as early as one day following the outburst.  ToO
observations continued for roughly the next month before regular monitoring
observations resumed.  Here, we report on observations of the persistent and
pulsed emission from the AXP leading up to, during and following the 2002
outburst of this source.  A detailed analysis of the burst properties is
presented in a companion paper (Gavriil et al.\ 2003).



\begin{figure}[!htb]
\centerline{
\psfig{file=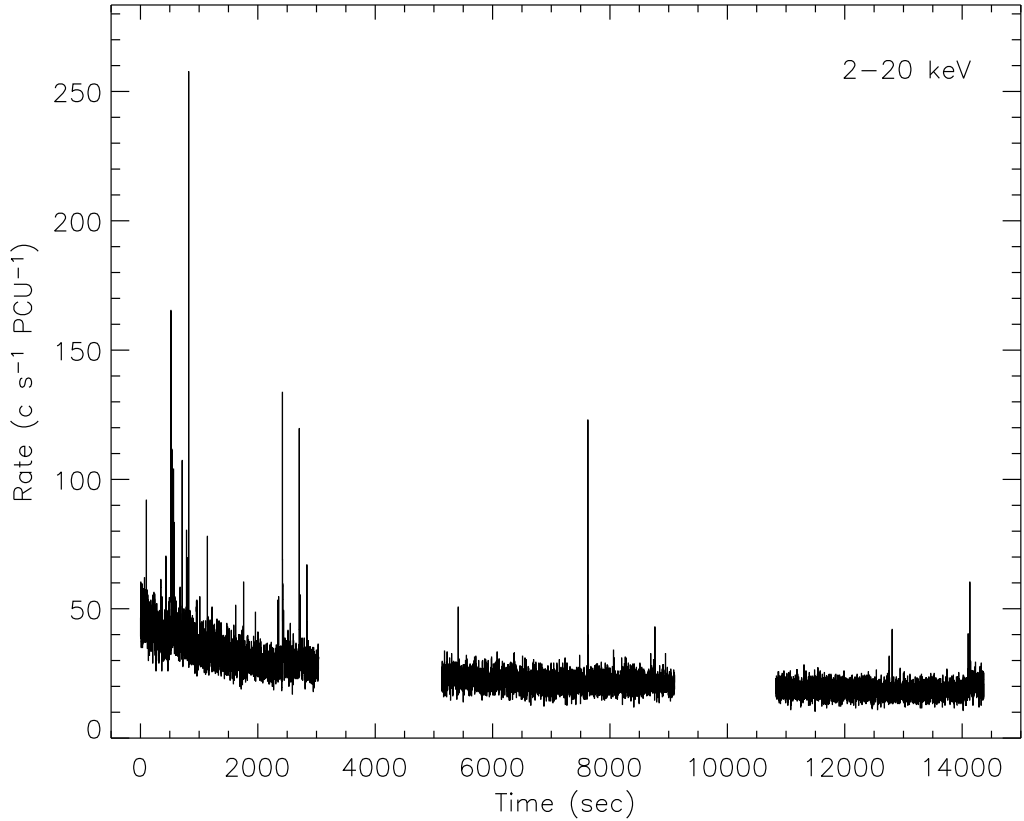,height=4.8in}}
\vspace{-0.05in}

\caption{{\it RXTE} PCA light curve of 1E~2259$+$586 recorded on 2002 June 18. 
The light curve displays counts within 2$-$20 keV at 1-s time resolution.}

\vspace{11pt}
\end{figure}

The PCA instrument aboard {\it RXTE} is a collimated (1 deg FWHM FOV)
proportional counter containing a mixture of xenon and methane gas (Jahoda et
al.\ 1996).  It is sensitive to X-rays in the 2$-$60 keV bandpass and has a
maximum effective area of $\sim$6500 cm$^{2}$ at 7 keV.  The data from the
PCA are read out in a number of different data acquisition modes.  Different
data modes were used depending upon the analysis performed.  In all of the
analysis, integration times including detected burst emission were eliminated
from the accumulated spectra and/or light curves.

With the exception of the first two ToO pointings directly following the
outburst, GoodXenonWithPropane data were acquired during each observation. 
These data were used for the pulse timing analysis described below.  For the
first two ToO observations, data modes better suited for studying bright bursts
were employed, however, no further burst activity was seen.  For these
pointings, the event data mode (E\_125us\_64M\_0\_1s) was used for the timing
analysis. 

Starting from {\it RXTE} production-level data, GoodXenonWithPropane and event
mode data were energy selected (2$-$10 keV) for all Xenon layers and binned
into light curves having 0.0625 s time resolution.  The time values in the
light curve were then corrected to the solar system barycenter.  For these
processing steps, the standard prescriptions for {\it RXTE} PCA data analysis
were
followed\footnote{http://heasarc.gsfc.nasa.gov/docs/xte/recipes/cook\_book.html}. 
The data were filtered further by eliminating times of high background and
bursts during the 2002 June 18 observation using custom software.  The count
rates were normalized to the number of PCUs on at a given time.

\subsection{Pulse Timing Analysis}

Within two days of the outburst, it was clear from the PCA data that a glitch
had occurred (Kaspi et al.\ 2003).  A new timing ephemeris was established and
refined with continuing PCA observations.  It gradually became apparent that
accompanying the sudden increase in frequency was a dramatic increase in
spin-down rate (i.e.\ torque) by a factor $\sim$2.  This torque excess decayed
rapidly over the next several weeks, approaching the pre-outburst level.  Here,
we extend the pulse timing ephemeris 15 months beyond the glitch, allowing us
to better characterize the frequency evolution post-glitch and quantify the
recovery time scale of the torque.

As with the {\it XMM-Newton} PN data, a phase folding technique was used to
determine the precise ephemeris from the {\it RXTE} PCA data.  Briefly, photon
arrival times, obtained using data in the energy range 2$-$10~keV were binned
with 62.5-ms time resolution, reduced to the solar system barycenter, then
folded at the nominal pulse period.  The folded pulse profiles for each
pointing were cross-correlated in the Fourier domain with a template pulse
profile obtained from pre-outburst observations only and a relative phase
(i.e.\ TOA) was measured.  Our analysis includes a total of 62 phase
measurements obtained between 2000 March and 2003 September, with 43
measurements obtained either during or post-outburst.

This particular analysis was complicated by the change in the pulse profile at
the time of the outburst (see \S3.4), as the cross-correlation procedure
assumes a fixed profile.  Gross pulse profile changes in which the relative
amplitudes of the two peaks were reversed resulted in a misidentification of
the standard fiducial point by the cross-correlator for a subset of the phase
measurements made during the outburst.  This was accounted for by aligning the
average pulse profile during the outburst observation with the template
profile.  The aligned outburst profile was then used as the template for phase
measurements during the outburst.  We verified this procedure by comparing the
phase measurements made using this method to those calculated using the
original template.  We found that where the cross-correlator identified the
appropriate peak in the original measurements, the revised measurements agreed
to within the errors.  We further checked that our method worked by verifying
that the results we describe below are insensitive to the omission of the phase
measurements made within 1 day of the burst activity where the pulse profile
changes were largest.

We attempted to fit the spin evolution through and following the glitch with
standard glitch models, that is, a simple jump in frequency, and a
two-component frequency jump in which one part decays exponentially (Eq.\ 2
below with $\Delta \nu_g = 0$). Neither provided a good fit to the phase data. 
Figure 6 shows the frequency evolution and phase residuals following
subtraction of the best-fit model including a sudden frequency jump and an
additional exponential decay. The best-fit parameters imply a total frequency
jump ($\Delta \nu$) of 6.3 $\times$ 10$^{-7}$ Hz, with a fraction $Q =$ 0.23
decaying on a time scale of $\sim$40 days.  However, as can be seen from the
Figure, the residuals from this model show a significant systematic trend at
the few percent level ($\chi^2$ = 447 for 54 dof).  Omitting the immediate
post-burst data in which the pulse profile had substantially changed does not
alter the result. It is possible, however, that the residuals are a result of
low-level systematic pulse morphology variations (see \S3.4).  This is hard to
rule out.



\begin{figure}[!htb]
\centerline{
\psfig{file=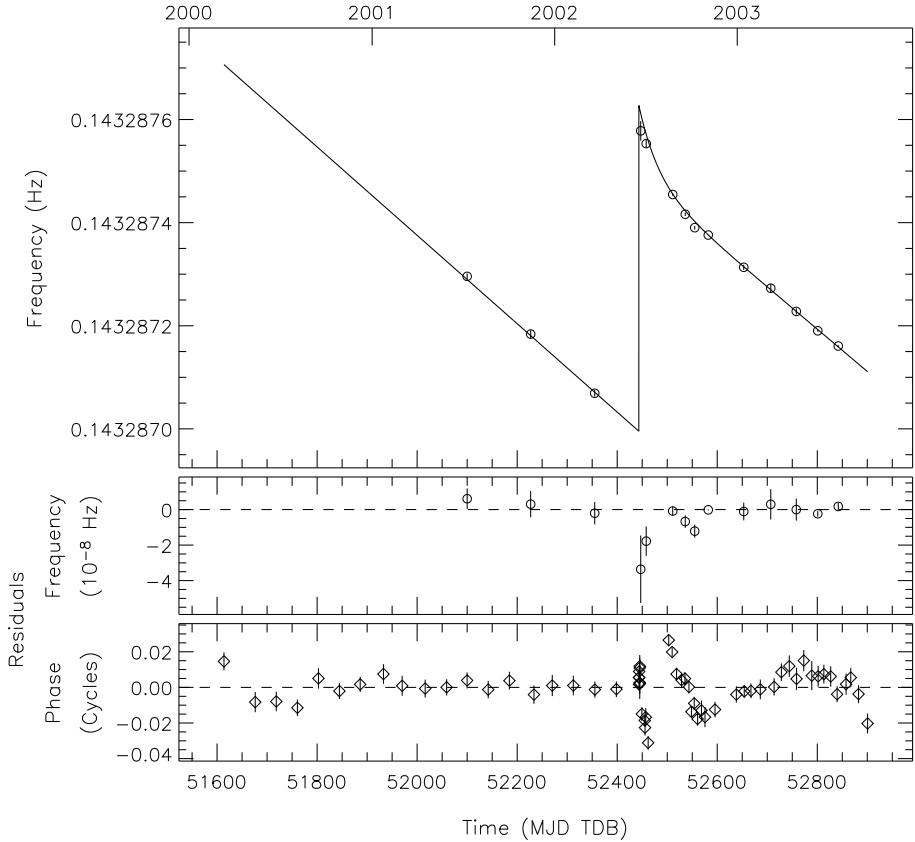,height=4.8in}}
\vspace{-0.05in}

\caption{{\it Top Panel} -- Frequency evolution of 1E~2259$+$586 around the
time of the outburst for a simple exponential recovery model.  See text for
details.   The solid line represents the best-fit model.  The circles denote
frequency measurements for independent subsets of data.  {\it Middle Panel} --
The frequency residuals of the independent frequency points minus the model.
{\it Bottom Panel} --  Phase residuals with respect to the best-fit model. 
Clearly, there is a systematic trend in the post-glitch residuals reflecting
the inadequacy of this model.}

\vspace{11pt}
\end{figure}

However, as we show next, a model in which a substantial portion of the glitch
is resolved in time provides a better fit to the data.  We employed a more
complex model previously invoked for fitting two glitches from the Crab pulsar
in which the rise in frequency is resolved (Lyne, Pritchard, \& Graham-Smith
1993; Wong, Backer \& Lyne 2001).  We model the frequency evolution as

\begin{equation}
\nu = \nu_0(t) + \Delta\nu + \Delta\nu_{g}~(1 - 
e^{-(t-t_g)/{\tau_{g}}}) - \Delta\nu_{d}~(1 - e^{-(t-t_g)/{\tau_d}}) + 
\Delta\dot{\nu}~t,
\end{equation}

\noindent
where $\nu_0(t)$ is the frequency evolution pre-glitch, $\Delta \nu$ is a
instantaneous frequency jump, $\Delta \nu_{g}$ is the resolved frequency jump
that grows exponentially on a time scale $\tau_{g}$, $\Delta \nu_d$ is the
post-glitch frequency drop that decays exponentially on a time scale $\tau_d$,
$t_g$ is the glitch epoch, and $\Delta \dot{\nu}$ is the post-glitch change in
the long-term frequency derivative.  

The full ensemble of phases was fit to the model above using a
Levenberg-Marquardt least squares fitting routine.  The fit improved
significantly over the simple jump in frequency and partial exponential
recovery ($\chi^2$ = 131 for 53 dof).  Due to the strong correlation between
the amplitudes of the two exponential factors ($\Delta\nu_{g}$ and
$\Delta\nu_d$) and similar associated time scales, our fit was mildly
non-convergent.  For this reason, we can only quote lower limits for each
amplitude.  Fixing either one of the exponential amplitudes to a value above
its lower limit allows the fit to converge.  The key parameter is the
difference between the two amplitudes which determines the peak frequency
following the glitch.  The best-fit parameters for this fit (including the
relationship between $\Delta\nu_{g}$ and $\Delta\nu_d$) are given in Table 4. 
The variance of the time scales ($\tau_g$ and $\tau_d$) is far less than the
amplitudes.  We quote the formal errors for these time scales using a fixed
$\Delta \nu_{g} =$ 2.3 $\times 10^{-6}$ Hz.  However, fixing $\Delta\nu_{g}$ to
the minimum allowed value yields significantly different time scales
($\tau_{g}$ = 12.8(7) days and $\tau_{d}$ = 17.4(6) days).  Going to higher
values for $\Delta \nu_{g}$ (and correspondingly $\Delta \nu_{d}$) does not
change the time scales significantly.


\begin{table}[!h]
\begin{minipage}{1.0\textwidth}
\begin{center}
\caption{Spin Parameters for 1E~2259$+$586 from 3.2 years of Phase-Coherent 
Timing using {\it RXTE} PCA data.} 
\vspace{10pt}
\begin{tabular}{lc} \hline \hline

Spin Frequency$^{a}$ , $\nu$ (Hz) & 0.14328703257(21) \\
Spin Frequency Derivative, $\dot{\nu}$ (Hz s$^{-1}$) & $-9.920(6) \times 10^{-15}$  \\
Epoch (MJD TDB) & 52400.0000 \\\hline
$\Delta \nu$ (Hz) & $5.25(12) \times 10^{-7}$ \\
$\Delta \nu_{g}{^b}$ (Hz) & $> 8.7 \times 10^{-7}$ \\
$\tau_{g}$ (days) & 14.1(7) \\
$\Delta \nu_d$ (Hz) & $\Delta \nu_g + \sim5 \times 10^{-9}$ \\
$\tau_d$ (days) & 15.9(6) \\
$\Delta \dot{\nu}$  (Hz s$^{-1}$) &  $+2.18(25) \times 10^{-16}$ \\
$t_g$ (MJD TDB) & 52443.13(9) \\\hline
RMS Timing Residual (ms) & 44.9 \\
Start Observing Epoch (MJD) & 51613 \\
End Observing Epoch (MJD) & 52900 \\

\hline\hline
\end{tabular}
\end{center}

\noindent$^{a}$ Numbers in parentheses represent 1$\sigma$ uncertainties
in the least significant digits quoted. \\
\noindent$^{b}$ Lower limit given at 90\% confidence. \\

\end{minipage}\hfill
\end{table}

The frequency evolution of 1E~2259$+$586 just before and following the outburst
as determined by our fit is shown in Figure 7.  The pre-outburst ephemeris is
fully consistent with the ephemeris determined through earlier monitoring of
this pulsar (Gavriil \& Kaspi 2002).  The glitch epoch ($t_g$) precedes the PCA
observation containing the burst activity by 12.5 $\pm$ 2.1 hours.  Note that
the glitch epoch precedes the observed burst activity whether the exponential
growth term is included in the fit or not (it is 4.7$\sigma$ in the case with
no growth term).  The exponential growth term clearly improves the fit to the
full data set; however, there is a systematic trend in the phase residual
cluster just post-glitch.  This is discussed in detail below.  Including the
growth term significantly reduces the time scale of the exponential decay term
to $\sim$16 days.  Finally, the long-term post-glitch spin-down rate {\it
decreases} significantly in magnitude (8$\sigma$) for both models.



\begin{figure}[!htb]
\centerline{
\psfig{file=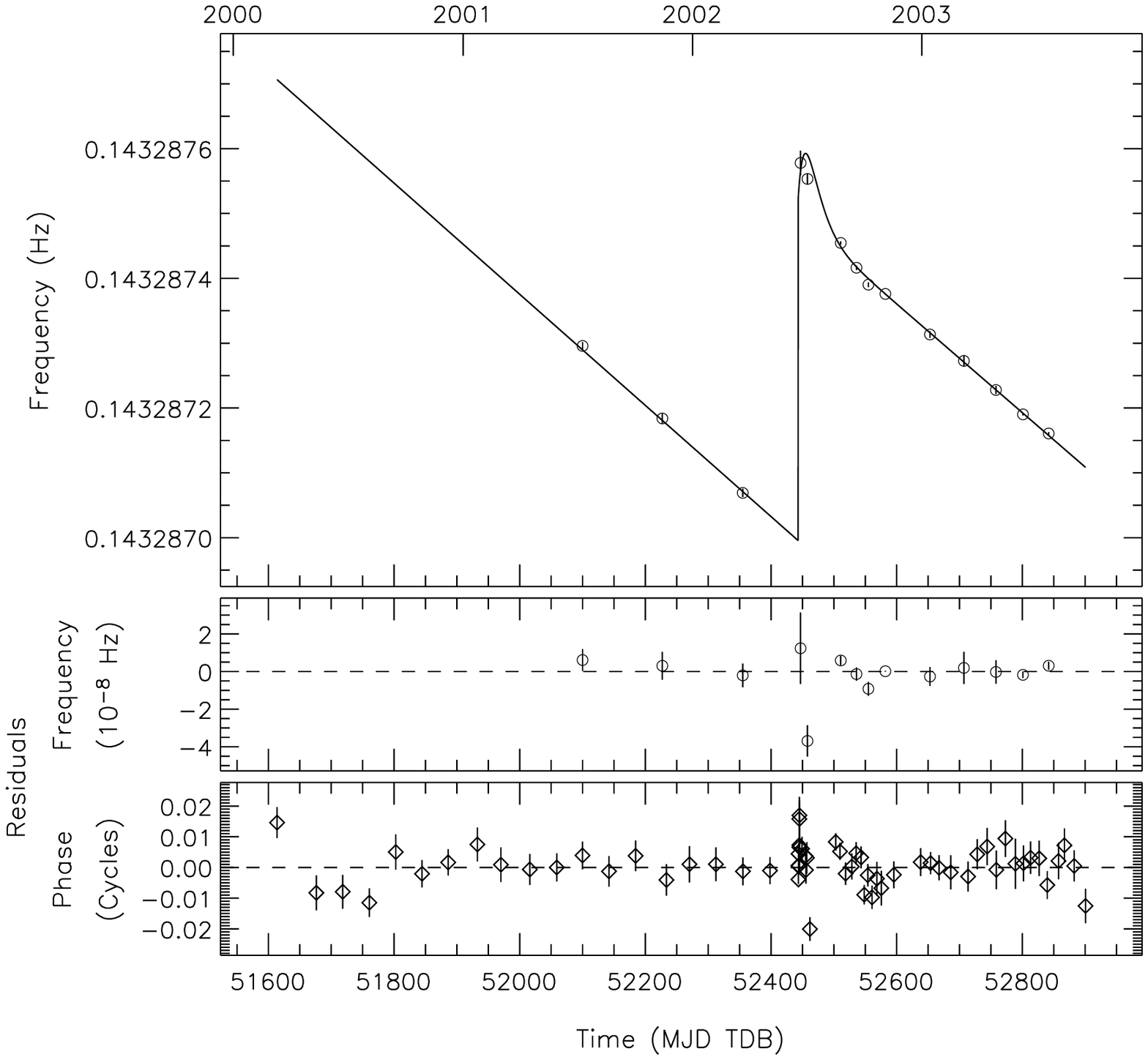,height=4.8in}}
\vspace{-0.05in}

\caption{{\it Top Panel} -- Frequency evolution of 1E~2259$+$586 around the
time of the outburst for a model including an exponential rise and fall in
frequency post-glitch.  See text for details.   The solid line represents the
best-fit model (see Table~4 for model parameters).  The circles denote
frequency measurements for independent subsets of data.  The effect of the
glitch is obvious, as is the partial recovery.  {\it Middle Panel} --  The
frequency residuals of the independent frequency points minus the model. {\it
Bottom Panel} --  Phase residuals with respect to the best-fit model. Closer
inspection of the residual cluster just following the outburst epoch reveals
that there is a low amplitude systematic trend.  This is discussed further in
the text.}

\vspace{11pt}
\end{figure}

The parameters in Table~4 show that the glitch consisted of a total fractional
frequency increase  $\Delta\nu_{\rm max}/\nu = (4.24 \pm  0.11) \times 10^{-6}$
where $\Delta\nu_{\rm max}$ is the maximum upward excursion in frequency
relative to the pre-glitch ephemeris.  Of this frequency jump, a fraction $Q
\equiv (\Delta\nu_{\rm max} - \Delta\nu - \Delta\nu_{g} +
\Delta\nu_d)/\Delta\nu_{\rm max} = 0.185 \pm 0.010$ recovered within two months
following the glitch. 

We note that a decrease in the magnitude of the spin-down rate would be unique
among all known pulsar and AXP glitches (Shemar \& Lyne 1996; Kaspi \& Gavriil
2003).  This is discussed more in \S4.2.5.  Alternatively, some radio pulsars
have shown a long-term exponential rise in frequency post-glitch.  However, our
best fit to this model was excluded by the frequency data.

The exponential rise term in this frequency model is clearly preferred by the
data, however, several important caveats must be stated.  The exponential form
was chosen due to its success in modeling radio pulsar glitches (Lyne et al.\
1993).  However, for the 19 days following the glitch, there is a significant
deviation from this model that constitutes a large portion of the remaining
$\chi^2$ in the fit.  During this 19 day interval, the best fit model predicts
rapid spin up for the first $\sim$10 days after which the frequency derivative
again becomes negative.  If we fit only the data from the 12 days following the
glitch, we measure a frequency derivative of $-$1.0 $\pm$ 0.3 $\times 10^{-13}$
Hz s$^{-1}$ whereas the model predicts an average frequency derivative of $+$5
$\times 10^{-14}$ Hz s$^{-1}$.  Summing the frequency derivative measurement
error and the model error in quadrature, we find that the two values differ at
the $\sim5\sigma$ level.  Hence, there is no direct proof that there was
significant spin up during the first 12 days following the glitch.  Note also
that the frequency derivative measurement is only at the 3$\sigma$ level, so
some spin up cannot be completely ruled out.  Another caveat when considering
the exponential rise in frequency is the pulse profile was undergoing large
changes during this same time interval (see \S3.4), thus comprimising our
ability to phase align with our template pulse profile.  In particular, if the
pulse profile was shifting in phase in a smooth manner as it changed shape,
then this shift in phase would manifest itself as an apparent change in
frequency.  It is not likely that the deviation from a pure exponential
recovery can be attributed entirely to a systematic shift in the pulse profile,
as that would require a large drift of $\sim$0.35 cycles within the two weeks
following the glitch.  Furthermore, even if we exclude from our fit the data in
which the pulse profile changes were largest, the growth term still
significantly improves the fit over the simple exponential recovery ($\Delta
\chi^2 = 136$ for 2 fewer dof).  We conclude from our analysis that there is a
significant deviation from a simple exponential recovery, however, due to gaps
in the data coverage (in particular from 19 to 60 days after the glitch) and
the inherent pulse profile changes, we could not precisely identify the manner
in which the frequency evolution deviated.

\subsection{Pulsed Flux History}

Coincident with the burst activity on 2002 June 18 was a sudden increase in the
pulsed flux from 1E~2259$+$586 (Kaspi et al.\ 2003).  The pulsed intensity of
the AXP decreased through the burst observation as did the burst rate, thereby
confirming the AXP as the origin of the burst emission.  Here, we put the
pulsed flux enhancement in context with the long-term pulsed flux history and
track the recovery back toward the pre-outburst level.

Using the ephemeris determined in the last section, pulse profiles of the PCA
data (2$-$10 keV) from 2000 March through 2003 June were constructed.  The
pulse profiles were generated for each pointing at times before and up to one
year following the outburst.  For the 2002 June 18 observation, the bursts were
removed and the data were split into $\sim$200 s segments before folding. 
Between 1 day and 19 days after the outburst, the data were grouped by
spacecraft orbit ($\sim$3 ks).  Once the pulse profiles were constructed, the
pulsed intensity was measured by first decomposing each pulse profile in terms
of its Fourier powers.  The power in the first 7 harmonics was summed to give
the RMS pulsed intensity (see Eq.\ 1).  The pulsed flux history of
1E~2259$+$586 is shown in Figure 8.  The large spike indicates the time of the
outburst.  Note that the pulsed flux has not yet returned to the pre-outburst
level.  The time scale and functional form of the recovery is studied in more
detail in \S3.6.



\begin{figure}[!htb]
\centerline{
\psfig{file=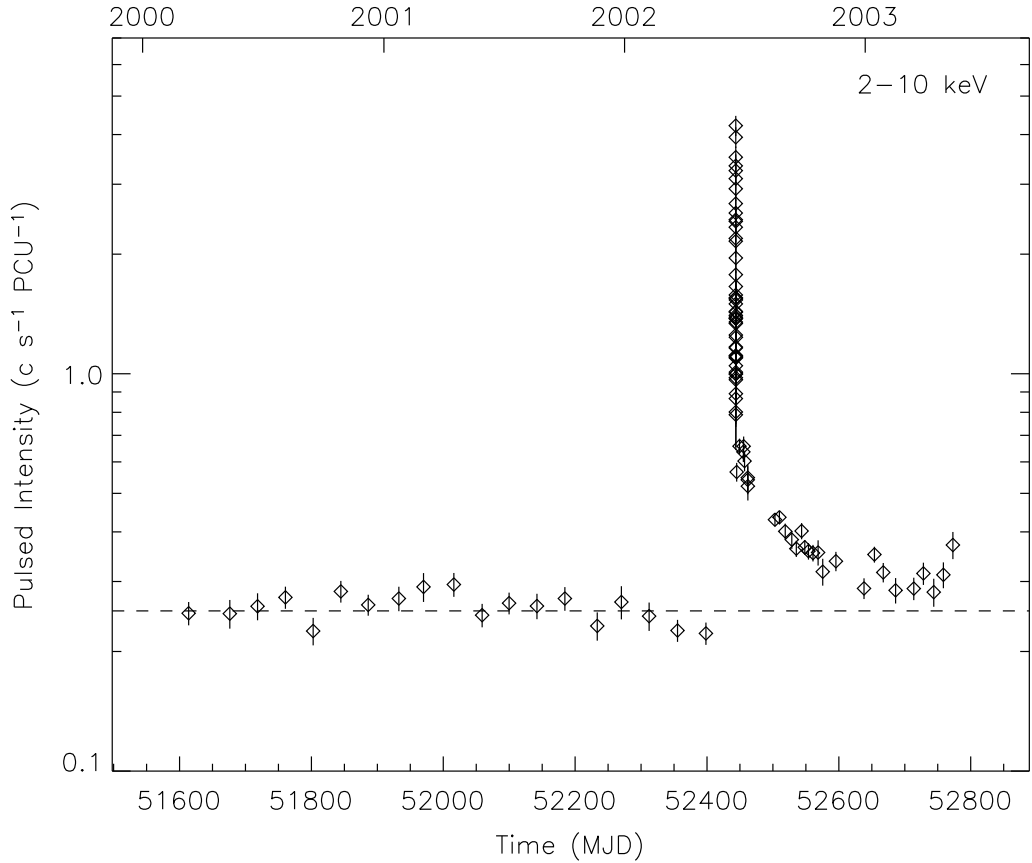,height=4.8in}}
\vspace{-0.05in}

\caption{The pulsed flux history of 1E~2259$+$586 (2$-$10 keV) from 2000 March
through 2003 June as measured using the {\it RXTE} PCA.  The sharp rise in the
pulsed intensity coincides with the burst activity.}

\vspace{11pt}
\end{figure}

\subsection{Pulsed Fraction}

The PCA is a sensitive X-ray detector with a wide FOV and no imaging
capabilities (for a fixed pointing).  Due to the design of the instrument, it
is not possible to reliably measure the pulsed fraction and/or spectrum of weak
X-ray sources such as 1E~2259$+$586 on account of uncertainties in the
background, particularly in the direction of the Galactic plane.  However,
during episodes where the AXP flux increases substantially, one can measure
accurately the pulsed fraction and spectrum (see \S3.5) with the PCA under
certain assumptions (Lenters et al.\ 2003).

The observed count rate in the PCA during pointings of 1E~2259$+$586 consists
of several components, namely the central source, instrumental background,
near-Earth background due to precipitating particles, the ``diffuse'' cosmic
background, the Galactic ridge, the bright SNR surrounding the AXP, and other
dim sources in the FOV.  To disentangle the spectrum of the central source from
all the other contributions to the net count rate, the remaining components
must somehow be modeled.  The instrumental and near-Earth background (i.e.\
local background) can be subtracted using housekeeping data and models provided
by the {\it RXTE}
team\footnote{ftp://legacy.gsfc.nasa.gov/xte/calib\_data/pca\_bkgd/}. 
Fortunately, the remaining components do not vary greatly on time scales of
months to years.

To estimate the count rate in the PCA of the remaining background components,
we did the following.  Using the last PCA observation (2002 May 04) taken
before the outburst, we selected counts in the range 2$-$10 keV from all PCUs
other than PCU0 and subtracted the background as estimated using the {\bf
FTOOL} {\tt PCABACKEST} with the combined model (CM) version of 2003 March 30. 
Note that PCU0 was excluded due to the loss of the propane layer and the
resulting large increase in the magnitude and variance of the background rate. 
Next, we measured the pulsed count rate (per PCU) during this observation. 
Assuming that the 2$-$10 keV pulse fraction measured during the {\it
XMM-Newton} observation from 2002 June 11 (Table 2) was the same on 2002 May
04, we determined the background rate in the PCA.  We use this ``cosmic''
background rate for all observations during and following the outburst where we
measure the pulsed fraction.  In using this count rate for our background, we
assume that ($i$) the AXP pulse fraction was the same on 2002 May 04 as it was
on 2002 June 11 and ($ii$) the cosmic background remains constant from 2002 May
04 until 2002 July 06 (the time of the last pulsed fraction measurement
reported here).  

Using this technique, we determined that the pulsed fraction of 1E~2259$+$586
decreased relative to the pre-outburst level while the source was burst active
(Kaspi et al.\ 2003).  Here, we have extended this analysis to later PCA
observations through 2002 July 06.  As with the {\it XMM-Newton} data, we
measured the pulsed fraction by subtracting the background and decomposing the
folded pulse profile in terms of its Fourier powers.  The RMS pulsed fraction
was determined from the sum of the first 7 harmonics using the formalism
described in \S2.2.  The {\it RXTE} PCA pulsed fraction measurements are
plotted along with the {\it XMM-Newton} measurements (Table 2) in Figure 9. 
We found that the 2$-$10 keV pulsed fraction decreased initially to a level of
$\sim$15\% where it remained for at least one day before rapidly returning to
the pre-outburst value within $\sim$6 days of the outburst.



\begin{figure}[!htb]
\centerline{
\psfig{file=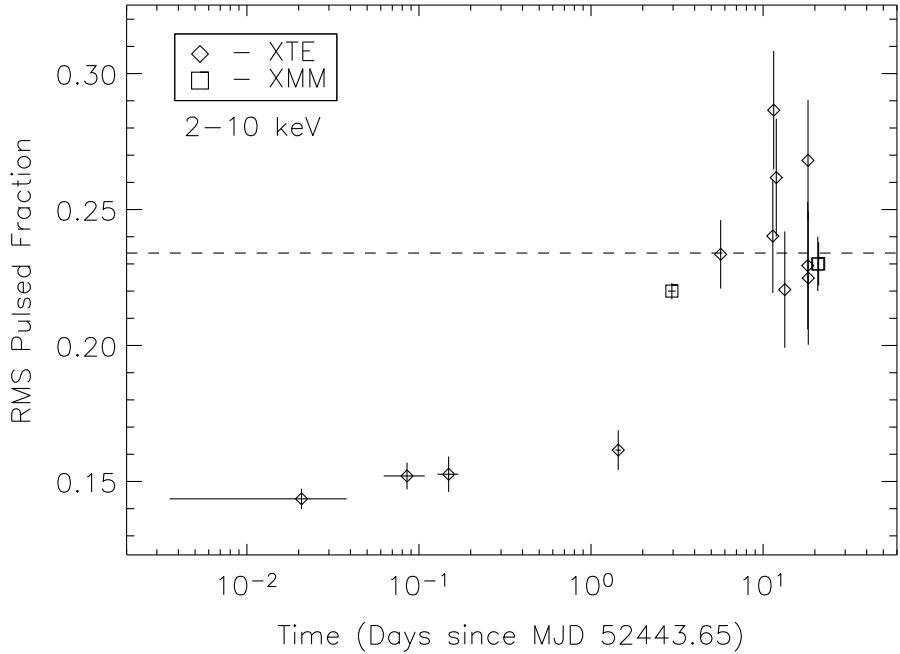,height=3.6in}}
\vspace{0.05in}

\caption{The time evolution of the 2$-$10 keV pulsed fraction of 1E~2259$+$586
through the outburst of 2002.  Diamonds denote measurements made with the {\it
RXTE} PCA and squares mark {\it XMM-Newton} measurements (Table 3).  The
horizontal dashed line marks the pulsed fraction from the {\it XMM-Newton}
observation one week prior to the outburst (Obs2).}

\vspace{11pt}
\end{figure}

\subsection{Pulse Profile}

Coincident with the detected burst activity in the PCA from 1E~2259$+$586 was a
sudden change in the folded pulse profile (Kaspi et al.\ 2003).  In the energy
range 2$-$5 keV, the relative amplitudes of the two peaks were switched while
the source was burst active.  In this energy range, the pulse profile returned
to near its pre-outburst form within $\sim$6 days.  No changes like this have
been seen in more than 6 years of monthly monitoring with the PCA (Gavriil \&
Kaspi 2002), although similar changes have been noted prior to the PCA
monitoring (Iwasawa, Koyama \& Halpern 1992).  Here, we investigate further the
pulse profile evolution over a longer time baseline and as a function of
energy.

The temporal evolution of the folded pulse profile of 1E~2259$+$586 from 2$-$10
keV is shown in Figure 10.  From the pulsed flux analysis, we know that the
amplitudes of both peaks increased during the 2002 June 18 observation relative
to their pre-outburst amplitudes.  Furthermore, the pulsed flux decreased
significantly during the burst observation ({\it RXTE} orbits 1$-$3).  This
allows us to conclude that the amplitude of peak 1 is decaying more rapidly
than peak 2 during the burst observation (as opposed to peak 2 growing relative
to peak 1).  The variability in the 18 days following the burst activity
(panels 5$-$8) shows erratic behavior in the pulse morphology.  It is not until
several weeks after the outburst that the pulse profile closely resembles its
pre-outburst form within this energy range.  Even at several months following
the outburst the differences are significant.  Specifically, the bridge of
emission connecting the two peaks is higher than it was pre-outburst.


\begin{figure}[!p]
\centerline{
\psfig{file=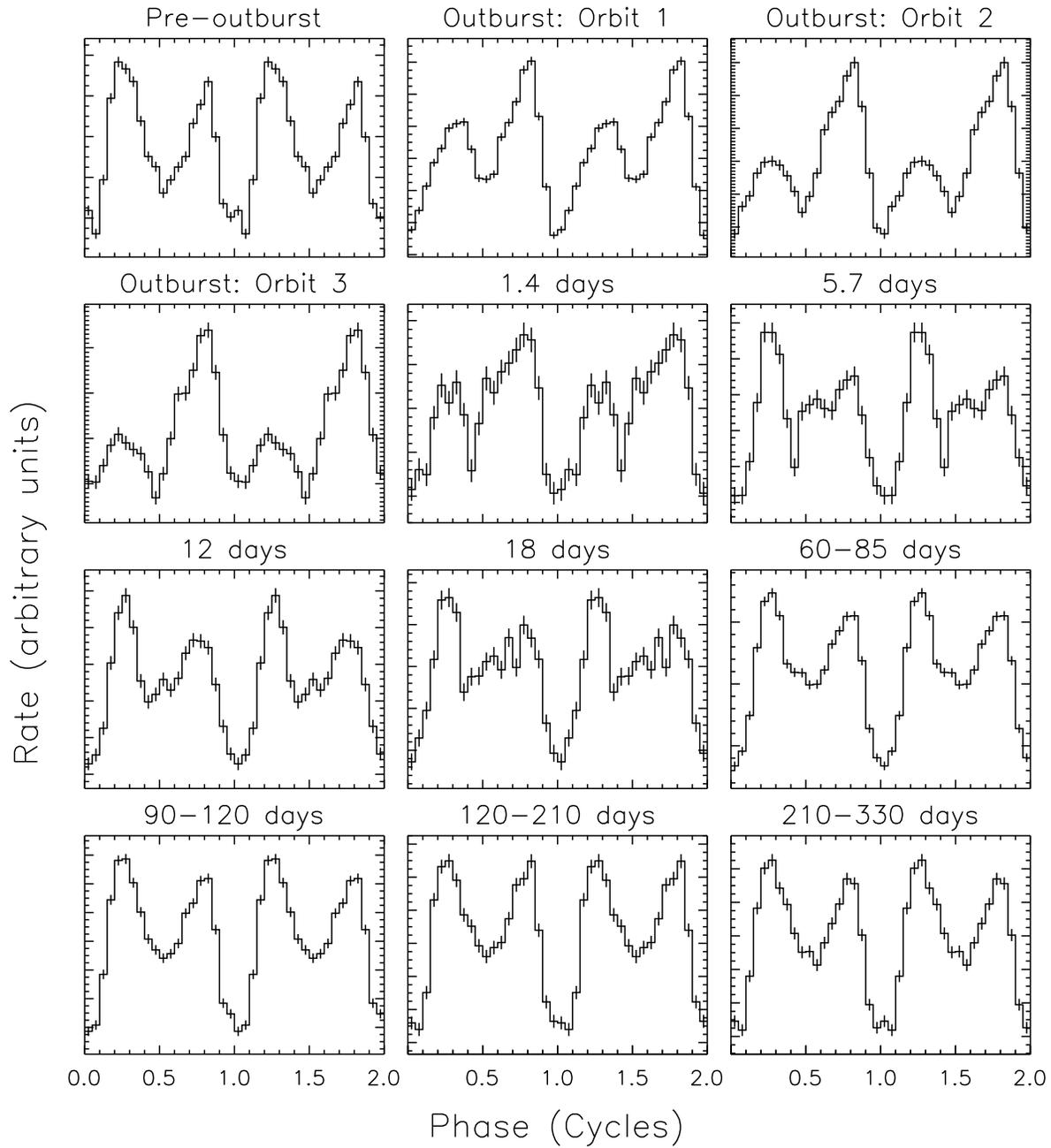,height=7.0in}}
\vspace{0.05in}

\caption{The time evolution of the 2$-$10 keV pulse profile of 1E~2259$+$586
through the outburst of 2002.  Profiles have arbitrary flux units and two pulse
cycles are shown for clarity.  Time increases from left to right and top to
bottom.}

\vspace{11pt}
\end{figure}

We quantified the change in pulse shape (2$-$10 keV) by decomposing the pulse
profile in terms of its Fourier powers following Equation 1.  As is clear from
Figure 11, the ratio of the power in the second harmonic to the first harmonic
(i.e.\ fundamental frequency) is the dominant factor governing the pulse shape
change.  During quiescence, the second harmonic contains more than 80\% of the
total power.  During the observation containing the bursts, the power in the
first harmonic increased such that there is actually as much or more power in
this harmonic compared with the second harmonic.  The ratio of the power in the
first two harmonics recovers gradually over the next several months.


\begin{figure}[!p]
\centerline{
\psfig{file=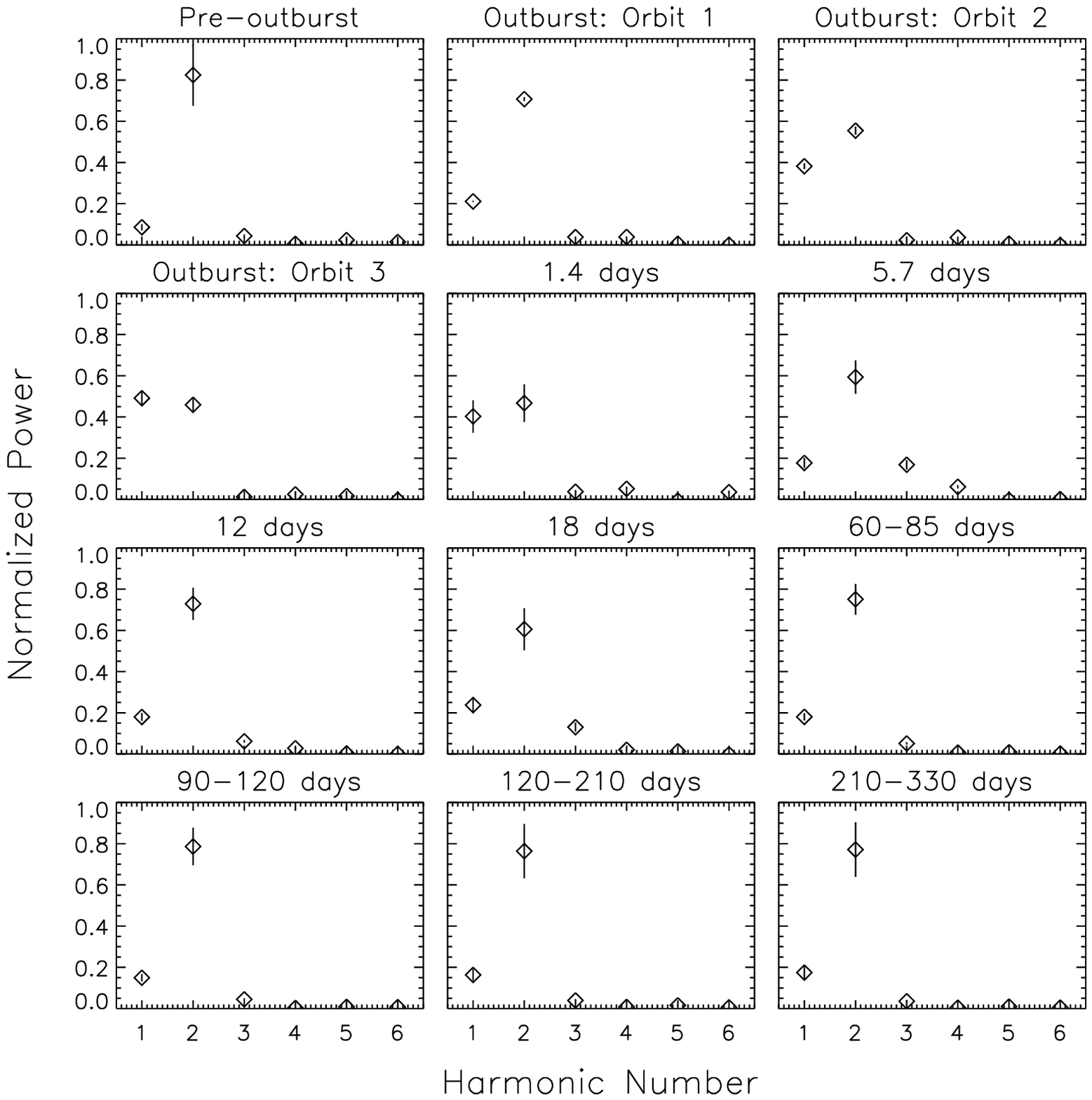,height=7.0in}}
\vspace{0.05in}

\caption{The time evolution of the Fourier power harmonic distribution of the
2$-$10 keV pulse profile of 1E~2259$+$586 through the outburst of 2002.  Power
levels have been normalized to the total power of the first 7 harmonics. Time
increases from left to right and top to bottom.}

\vspace{11pt}
\end{figure}

Lastly, we investigated the dependence of the pulse shape on photon energy
during the burst observation when the pulse shape was distorted the greatest. 
We find that there is a significant energy dependence of the pulse shape that
becomes more prominent toward the end of the burst observation.  Specifically,
peak 2 becomes narrower at high energies and the centroid of this peak lags in
phase by $\sim$0.02 cycles (0.15 s) from 2 to 15 keV.  Both the centroid and
phase of peak 1 are consistent with remaining unchanged versus energy. 
Similarly, the relative amplitudes of peak 1 and peak 2 are consistent with
being constant in this energy range.

\subsection{Spectroscopy}

During the outburst detected on 2002 June 18, the X-ray spectrum of
1E~2259$+$586 became much harder (Kaspi et al.\ 2003).  Initially, the
blackbody temperature increased to $\sim$1.7 keV (from $\sim$0.42 keV) and
the photon index flattened to $\sim$2.5 (from $\sim$4.2).  During the next
4 hours, the spectrum of 1E~2259$+$586 evolved back toward its quiescent shape,
but did not recover fully.  Here, we analyze the {\it RXTE} PCA observations
that took place over the next few weeks to track the recovery of the AXP
spectrum.

As with the pulsed fraction, measuring the X-ray spectrum of 1E~2259$+$586 with
the {\it RXTE} PCA is not straightforward, so we used a similar technique in
subtracting the background as that employed to measure the pulse fraction (see
\S3.3).  As before, the last PCA observation carried out on 2002 May 04, 46
days before the outburst, was used to estimate the cosmic background in the PCA
FOV. First, the local background during this observation was estimated using
the {\bf FTOOL} {\tt PCABACKEST} and the combined model of 2003 March 30.  The
observed Standard 2 spectrum\footnote{As with the pulsed fraction analysis, all
data from PCU0 were excluded from the spectral analysis.} was grouped such that
there were at least 25 total counts per bin (source plus background) and then
fit (less the local background) using {\bf XSPEC} v11.2.0 to a multi-component
model.  The model consisted of the standard AXP spectrum plus another blackbody
and a Gaussian line.  The second blackbody component and Gaussian line were
each modified by interstellar absorption with a fixed column density ($N_{H}$ =
2 $\times$ 10$^{22}$ cm$^{-2}$).  The AXP spectral parameters were frozen to
the values obtained from the {\it XMM-Newton} observation of 2002 June 11 (see
Table 3), 37 days following this observation and 7 days prior to the outburst. 
Using this model, we obtained a good fit to the PCA data between 2.5 and 20.0
keV ($\chi^2_{\nu}$ = 0.71 for 35 dof).  The free components in this fit (i.e.\
blackbody and Gaussian line), therefore, define the spectrum of the remaining
background in the PCA for this particular pointing at this particular epoch. 
This model was used to define the cosmic background in the PCA detector during
and directly following the burst activity assuming that these components (e.g.\
Galactic ridge, SNR, etc.) remain constant between 2002 May 04 and 2002 July
06.

A standard 2 spectrum was accumulated for the 2002 June 18 observation when the
bursts were detected.  Note that the bursts themselves were eliminated from the
data used to create the spectrum.  No single-component model provided an
adequate fit to the data.  For example, a fit to a power-law model yielded a
statistically unacceptable  $\chi^2$ of 384.7 for 37 degrees of freedom. 
Conversely, the standard blackbody plus power law model resulted in a very good
fit ($\chi^2$ = 35.8 for 35 degrees of freedom).  We measure a significantly
harder spectrum than ever before seen in this AXP with $kT$ = 1.23 $\pm$ 0.4
and $\Gamma$ = 2.83 $\pm$0.07.

Next, we split the 2002 June 18 into seven separate segments to search for time
evolution of the spectrum.  Spectra were also accumulated for each PCA
observation from 2002 June 20 through July 06.  We fit the blackbody plus power
law model to each of these PCA observations (2.5$-$20.0 keV).  The {\it RXTE}
fit results are shown in Figure 12 along with the {\it XMM-Newton} measurements
given in Table 3.



\begin{figure}[!htb]
\centerline{
\psfig{file=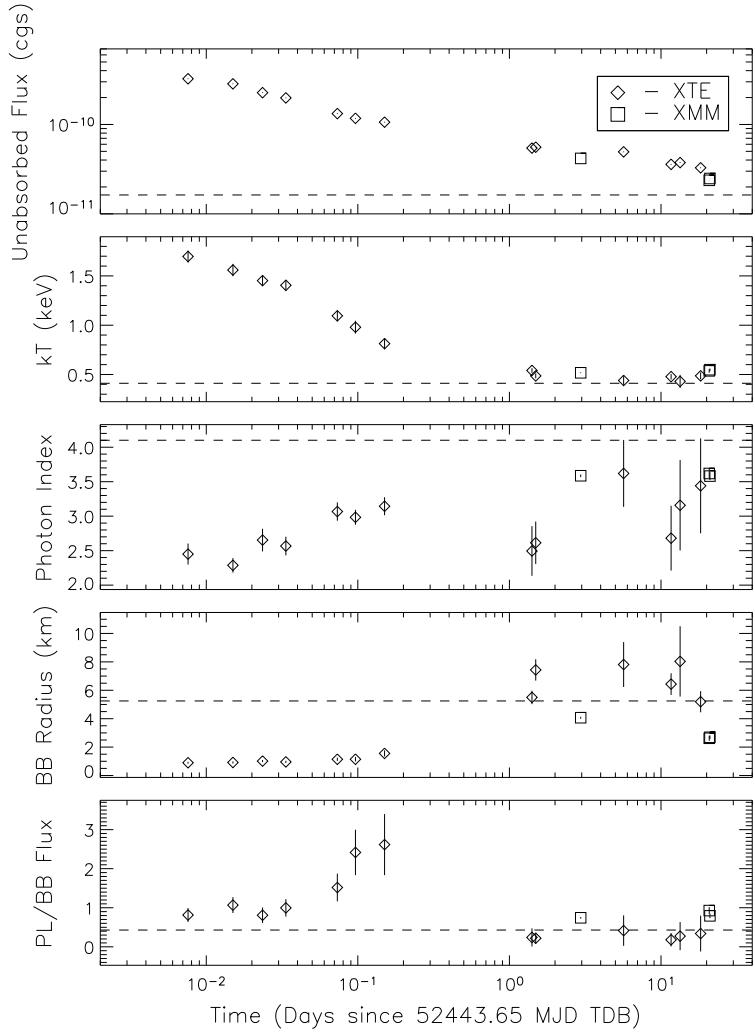,height=6.0in}}
\vspace{-0.05in}

\caption{The spectral evolution of 1E~2259$+$586 through and following the
outburst of 2002.  From the top panel down: the unabsorbed flux (2$-$10 keV),
the blackbody temperature ($kT$), the photon index, the blackbody radius, and
the ratio of power-law (2$-$10 keV) to bolometric blackbody flux.  A distance
of 3 kpc was assumed (Kothes, Uyaniker, \& Aylin 2002) to compute the blackbody
radius.  Horizontal dashed lines denote the values of each parameter during the
{\it XMM-Newton} observation one week prior to the outburst.}

\vspace{11pt}
\end{figure}

We tested the robustness of our technique by varying the assumptions that the
spectrum of 1E~2259$+$586 remains constant during the 44 days leading up to the
burst activity.  Specifically, when fitting for the PCA cosmic background in
the 2002 May 04 observation, we tested a range of values for the spectral
parameters ($kT$ and $\Gamma$) of 1E~2259$+$586 within their respective
historical ranges (see \S2.3).  Obviously, this did alter the cosmic background
fit parameters, but resulted in only insignificant changes ($\lesssim$1\%) to
the measured spectral parameters of the AXP for the 2002 June 18 observation. 
For the final PCA observation fit in this analysis where the AXP was dimmest
(2002 July 06), the relative change in these parameters is larger ($\sim$5\%),
but still less than 1$\sigma$.

The unabsorbed flux decayed rapidly within the first day of the outburst.  The
functional form of the decay depends critically upon the reference epoch
chosen.  Both power-law and exponential decay models yield acceptable fits to
the data.  At 20 days following the outburst, the AXP is still a factor of
$\sim$2 brighter than its nominal (quiescent) flux level.  The flux decay is
covered in greater detail in the next section.  The remaining spectral
parameters shown here recover to within 25\% of their pre-burst levels within
the first $\sim1-3$ days.  Using the pulsed flux history as an indicator for
the recovery time scale, a full recovery of the spectral parameters would not
be expected until $\sim$1 year later.  However, variation in these spectral
parameters at the $\sim$25\% level has been seen outside of burst activity (see
\S2.3), thus these parameters may not decay further.

It is interesting to note that the X-ray spectrum of 1E~2259$+$586 was harder
when the spin-down rate was higher during the first month following the
glitch.  This behavior in 1E~2259$+$586 is qualitatively consistent with the
spectral hardness versus spin-down rate correlation found by Marsden \& White
(2001) who considered the AXPs and SGRs as a whole.  This suggests that
individual members of this class that have shown significant variability in
spin-down rate may also show correlated spectral variations.

We note that the later {\it RXTE} PCA spectral measurements ($>$1 day) agree
reasonably well with the {\it XMM-Newton} measurements contained within this
interval, however, they are systematically offset from one another.  This could
be due to systematic effects in our background subtraction method, a deviation
in the power-law spectral model at high energies where the two instrumental
responses do not overlap (12$-$20 keV), and/or a systematic offset in the
cross-calibration between the {\it XMM-Newton} PN and the {\it RXTE} PCA.

\subsection{Flux Decay and Energetics}

To better quantify the flux decay of 1E~2259$+$586 following the outburst of
2002 June, we combined the {\it XMM-Newton} flux measurements with our {\it
RXTE} PCA pulsed flux measurements which span a much broader temporal
baseline.  The advantage of using the PCA pulsed flux measurements as opposed
to the phase-averaged flux measurements is that the systematic errors present
in the background subtraction are not a concern.

The pulsed flux measurements were converted to unabsorbed phase-averaged fluxes
in the following way.  First, we determined a conversion factor between the two
by calculating the ratio of the pre-outburst unabsorbed flux measured with {\it
XMM-Newton} to the average PCA pulsed flux for the year leading up to the
outburst.  Assuming that the pulsed fraction and spectrum do not change, this
factor can be multiplied with subsequent PCA pulsed flux values to estimate the
unabsorbed phase-averaged flux at those times.  However, we know that both the
pulsed fraction {\it and} the energy spectrum changed during this outburst.  We
corrected for the brief period where the pulsed fraction decreased by
multiplying by an additional factor of the ratio of the nominal pulsed fraction
to the observed pulsed fraction at those times.  This factor was only applied
to the PCA pulsed flux measurements within the first 2 days following the burst
activity where the fraction dropped from 23\% to $\sim$15\%.  Computing this
conversion factor for a broad energy range (2$-$10 keV) reduces the effects of
spectral changes.  In fact, when we applied our pulsed flux to unabsorbed
phase-averaged flux conversion factor to the data where we have independent
unabsorbed flux measurements at early times in the outburst ($<$20 days after
burst activity), the agreement between the two measurements was found to be
quite good (Figure 13).  Since the spectral differences are greatest at these
times in the outburst, we feel that this technique is a robust one for
constructing a long-term light curve for the source.



\begin{figure}[!htb]
\centerline{
\psfig{file=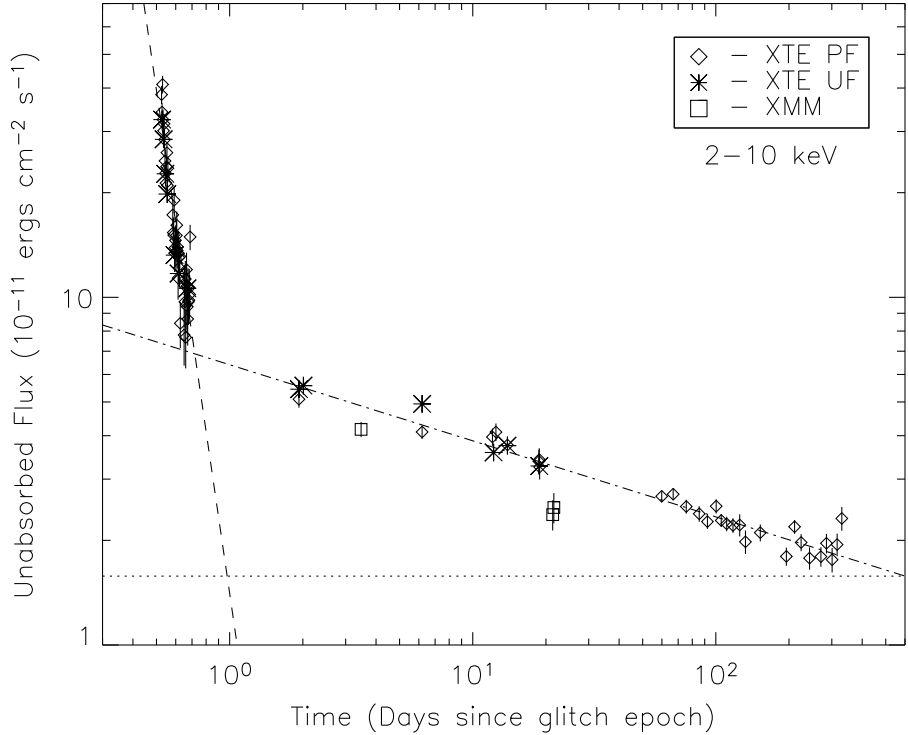,height=4.2in}}
\vspace{-0.05in}

\caption{The time evolution of the unabsorbed flux from 1E~2259$+$586 following
the 2002 June outburst.  The glitch epoch (Table 5) is used as the reference
time for this plot.  Diamonds denote inferred unabsorbed flux values calculated
from {\it RXTE} PCA pulsed flux measurements.  Asterisks and squares mark
independent phase-averaged unabsorbed flux values from {\it RXTE} and {\it
XMM-Newton}, respectively.  The dotted line denotes the flux level measured
using {\it XMM-Newton} 1 week prior to the glitch.  The dashed line is a
power-law fit to the PCA flux measurements during the observations containing
the burst activity ($<$1 day). The dot-dash line marks the power-law fit to all
data $>$1 day following the glitch.  See text for further details.}

\vspace{11pt}
\end{figure}

In this way, we have taken the pulsed flux measurements presented in Figure 8
and converted them to phase-averaged unabsorbed flux values.  These data are
plotted relative to the glitch epoch (Table 4) in Figure 13 in addition to the
three {\it XMM-Newton} flux measurements.  Clearly, the flux decay is not well
described by a single component model (e.g.\ exponential or power law).  The
temporal decay of the flux during the PCA observations containing the burst
activity ($<$1 day) is much more rapid than the decay during the year following
the burst activity.  We split the data into two segments ($<$1 day and $>$1 day
after the glitch), and fit each independently to a power-law model ($F \propto
t^{\alpha}$).  The measured temporal decay indices for the two segments are
$-$4.8 $\pm$ 0.5 and $-$0.22 $\pm$ 0.01, respectively.

The {\it RXTE} All-Sky Monitor (ASM) observed 1E~2259$+$586 at 03:50 and 14:43
UT on 2002 June 18, before the observed burst activity and following the
measured glitch epoch, although the first observation is within the errors of
the measured glitch epoch.  If we extrapolate the slower flux decay model
($\alpha$ = $-$0.22) back to the time of the ASM observations, we find that the
expected flux values (14 and 8 $\times$ 10$^{-11}$ ergs cm$^{-2}$ s$^{-1}$,
respectively) are comfortably below the 99\% confidence upper limit of 1
$\times$ 10$^{-9}$ ergs cm$^{-2}$ s$^{-1}$ (Kaspi et al.\ 2003), thus the ASM
limits are not constraining for this component.  For the steeper component
($\alpha$ = $-$4.8) containing the burst activity, we find that the expected
flux at the time of the first ASM observation is several orders of magnitude
above the upper limit.  The second ASM observation (52443.611 MJD TDB) is
unconstraining.  This suggests that the onset of burst activity (associated
with the first decay component) must have come after the earlier ASM
observation (52443.158 MJD TDB) or the flux enhancement deviated from this
steep decay between the glitch epoch and the {\it RXTE} PCA observations.  One
possibility is that the onset of this burst activity was delayed with respect
to the glitch epoch.  In fact, a much later onset time for the burst activity
is inferred from the time evolution of the burst recurrence frequency (Gavriil
et al.\ 2003).

The absence of a fiducial point for the initial, rapid flux decay associated
with the burst activity does not allow us to accurately measure the temporal
decay index of this component.  Choosing a reference epoch just before the time
of the first bursts detected with the PCA yields a decay index less than unity
(in magnitude).  Thus, we can only constrain the index of the flux decay within
the time span containing the burst activity to be $<$6.6 (in magnitude).  Note
that as the reference epoch approaches the start of the burst observation, the
flux decay becomes steeper than a power law in form (e.g.\ exponential).  This
is discussed further in \S4.1.1.  Unlike the early component, the temporal
index of the more gradual flux decay in the months following the glitch
($\alpha$ = $-$0.22) is insensitive to varying the reference epoch between the
time of the glitch and the beginning of the observed burst activity.

We next used these power-law fits for estimating the total energy released
during this outburst.  We integrated each model, less the quiescent flux level
(Figure 13), only over the time ranges where we have flux measurements (i.e.\
from the start time of the burst observation onward).  In spite of the large
range of allowed temporal decay indices, the energy we measure in the early
flux decay component is well determined since we integrate the model {\it only}
over the interval where we have observations (i.e.\ we do not extrapolate the
model back toward the glitch epoch).  Assuming a distance of 3 kpc to
1E~2259$+$586 (Kothes et al.\ 2002), we measure an energy release (2$-$10 keV)
of 2.7 $\times$ 10$^{39}$ ergs and 2.1 $\times$ 10$^{41}$ ergs for the fast and
slow decay intervals, respectively.  In terms of the overall energy budget, the
energy released in the bursts themselves (6 $\times$ 10$^{37}$ ergs over 2$-$60
keV [Gavriil et al.\ 2003]) is insignificant in comparison to the excess
persistent emission in X-rays released during the year following the outburst. 
Moreover, the excess persistent flux emitted during the interval containing the
burst activity is insignificant in comparison to the total energy released
during the year-long flux decay.

\section{Discussion}

Virtually all measurable X-ray properties of 1E~2259$+$586 changed suddenly and
dramatically during the 2002 June outburst.  Continued observations with {\it
RXTE} and {\it XMM-Newton} have allowed us to track the recovery of several
source parameters shown to change during this outburst (Kaspi et al.\ 2003) and
identify additional parameters that were similarly affected. Many of the
observed variations resemble phenomena seen in other classes of neutron stars,
namely SGRs and radio pulsars.  Here, we compare the AXP outburst properties
with similar phenomena seen in those source classes in the hopes of identifying
similarities and differences that can help elucidate the physical properties of
these different manifestations of young neutron stars.  Specifically, we
consider ($i$) the radiative properties of the persistent and pulsed emission
during and following the outburst and compare these with those seen in SGRs,
since that is the only other source class for which such outbursts have been
seen and ($ii$) the rotational behavior of the pulsar and compare it both with
behavior seen in radio pulsars as well as in SGRs and in another AXP, 1RXS
J1708$-$4009. Finally, we point out that the detection of low-intensity
outbursts in AXPs has important implications for our estimates of the number of
active magnetar candidates in our Galaxy.  The 1E~2259$+$586 burst properties
and their relation to SGR burst properties are considered separately (Gavriil
et al.\ 2003)

\subsection{Transient X-ray Emission and Pulse Properties}

We have shown that there are two components to the flux decay in 1E~2259$+$586
during the 2002 June outburst.  There is a rapid decay of the flux during the
observations containing the burst activity and then there is a more gradual
flux decay seen in the year following the glitch.  During the initial flux
decay, the spectrum was considerably harder than at all other times while the
spectral hardness at times $>$1 day after the glitch are consistent with
pre-glitch spectral measurements.  The spectral differences in the two decay
components (total energy, blackbody radius, etc.) point toward separate
physical mechanisms for the two flux enhancements (see Figures 9 and 12).

Coincident with the glitch and burst activity was a dramatic change in the
pulse profile of 1E~2259$+$586.  The majority of the observed shape change
recovered within $\sim$6 days, but there was still some residual change that
slowly decayed over the months following the outburst (\S3.4).  The 2$-$10 keV
pulsed fraction dropped to $\sim$15\% during the observed burst activity and
quickly recovered to the pre-outburst level of 23.4\% within $\sim$6 days
(\S3.3).

Here, we compare the observed properties of the flux enhancement and pulse
properties in 1E~2259$+$586 to qualitatively similar behavior detected in SGRs
and briefly discuss how the changes seen in 1E~2259$+$586 can be accomodated
within the magnetar model.

\subsubsection{Comparison to SGR Outbursts: X-ray Flux and Spectrum}

The richest SGR database with which to make an empirical comparison to the
1E~2259$+$586 outburst comes from the most active SGR during the last 6 years,
SGR~1900$+$14.  This SGR has been observed on 14 separate occasions since 1997
by imaging X-ray telescopes and more than 100 times with {\it RXTE}.  Within
this time span, SGR~1900$+$14 entered several burst active episodes; the most
notable of which was the outburst that began on 1998 August 27 with a giant
flare having a total energy $\sim$10$^{44}$ ergs (e.g.\ Feroci et al.\ 2001). 
Coincident with this giant flare was a large increase in the persistent and
pulsed flux from the source (e.g.\ Woods et al.\ 2001), in addition to a
dramatic change in the pulse profile and a timing anomaly (see \S4.2.3).

A spectral analysis of the {\it RXTE} PCA observation of SGR~1900$+$14 one day
following the giant flare shows that the blackbody temperature was higher than
the nominal temperature ($\sim$0.5 keV) at 0.94 keV (Woods 2003).  Two other
high fluence bursts from SGR~1900$+$14 have extended X-ray tails or afterglows
that show enhanced thermal emission at times reaching up to 4 keV (Ibrahim et
al.\ 2001; Lenters et al.\ 2003).  The thermal component of 1E~2259$+$586 shows
a similar brightening and temporal decay.  Here, the temperature rose from 0.4
keV up to 1.7 keV at the onset of the outburst before quickly decaying to 0.5
keV within the first few days.  Accompanying the temperature increase in
1E~2259$+$586 was a significant hardening of the photon index.  This is
different than what has been seen in SGR~1900$+$14 where there has been either
no change in photon index after some bursts (Lenters et al.\ 2003) or even a
slight {\it softening} of the non-thermal component of the spectrum (Woods
2003).

An analysis of four separate bursts/flares from SGR~1900$+$14 and their
associated afterglows (Lenters et al.\ 2003) shows that the emitted energy
within the afterglow (2$-$10 keV) corresponds to roughly 2\% of the burst
energy (25$-$100 keV).  Due to the bandpass of the PCA, we have only been able
to measure the burst energy within the 2$-$60 keV range (Gavriil et al.\
2003).  However, we can estimate the burst energy (25$-$100 keV) during the
2002 June outburst by taking the measured count fluence with the PCA (Gavriil
et al.\ 2003) and multiplying by the counts-to-energy conversion factor
determined for SGR~1900$+$14 ({G\"o\u{g}\"u\c{s}} et al.\ 1999).  This
conversion factor was used because the mean burst spectral hardness in the PCA
was similar for 1E~2259$+$586 (Gavriil et al.\ 2003) and SGR~1900$+$14
({G\"o\u{g}\"u\c{s}} et al.\ 2001).  The total energy was further modified by a
factor 2 increase to roughly compensate for gaps in the data due to Earth
occultation during the PCA pointing.  Assuming that the energy spectra of the
1E~2259$+$586 bursts above 60 keV is similar to SGR~1900$+$14, we estimated the
burst energy released by 1E~2259$+$586 to be $\sim$1 $\times$ 10$^{38}$ ergs
(25$-$100 keV).  The brightening of 1E~2259$+$586 during the 2002 June outburst
contained two components: a rapid decay during the initial burst observation
followed by a more gradual flux decay in the year following the glitch.  The
energy released from 1E~2259$+$586 within these two components is $\sim$30 and
$\sim$2000 times {\it greater} than the burst energy (Table 5) -- in stark
contrast to the value of 0.02 found for SGR~1900$+$14.  In fact, for a 2\%
afterglow-to-burst energy ratio, we can rule out the presence of an
intermediate burst from 1E~2259$+$586 preceding the 2002 June 18 PCA
observation.  Assuming that only the hard component of the decay constituted
the burst afterglow, the energy of the hypothetical burst powering this
afterglow must be $1.5\times 10^{41}$ ergs (equivalent fluence of $1.4 \times
10^{-4}$ ergs cm$^{-2}$).  The Konus detector aboard the {\it Wind} spacecraft
located at the L1 point between Earth and the Sun maintained continuous
coverage of 1E~2259$+$586 from the time of the glitch through the first PCA
observation and did not detect any emission from the AXP (S.~Golenetskii,
private communication).  The upper limits on the burst fluence (17$-$70 keV)
derived from the Konus data are $1.7 \times 10^{-7}$ ergs cm$^{-2}$ for bursts
of duration less than 3 s and $1.7 \times 10^{-7}$ ($\Delta t$/3 s)$^{1/2}$
ergs cm$^{-2}$, where $\Delta t$ is the burst duration, for bursts lasting
between 3 and $\sim$1000 s.  Clearly, the putative burst with fluence $1.4
\times 10^{-4}$ ergs cm$^{-2}$ can be excluded for durations less than
$\sim$1000 s.

The temporal decay of the flux from SGR~1900$+$14 following the August 27 flare
follows a power law in time with an exponent $-$0.71 during the 40 days
following the flare (Woods et al.\ 2001).  For a power-law fit to the flux
evolution during the burst activity, we can only constrain the decay to have an
exponent less than 6.6 (in magnitude) due to the uncertainty in the reference
epoch (see \S3.6).  Thus this decay component is not inconsistent with a $-$0.7
power law.  The more gradual flux decay in the months following the glitch
obeyed a power law in time that scaled as $t^{-0.22 \pm 0.01}$, significantly
flatter than the decay following the August 27 flare or any other flux decay
following bright bursts from SGR~1900$+$14 (Lenters et al.\ 2003; Feroci et
al.\ 2003).

We can also compare the energetics and flux decay of 1E~2259$+$586 to those
seen in SGR~1627$-$41.  This SGR has shown one outburst in 1998 where 98\% of
the emitted burst energy was concentrated into a narrow 3-week window (Woods et
al.\ 1999a).  The flux from the source decayed over the next 3 years
approximately as a power law in time with an exponent $-$0.47 (Kouveliotou et
al.\ 2003).  The last two {\it Chandra} observations of this SGR show that the
flux has leveled out (at least temporarily).  Since there are no pre-outburst
flux measurements to establish the ``quiescent'' flux level of this SGR, we
take the two latest flux measurements as the quiescent flux level.  Under these
assumptions, the burst tail or afterglow energy during the four years following
the 1998 outburst is comparable to the energy released in the bursts themselves
(Kouveliotou et al.\ 2003).


\begin{table}[!h]
\begin{minipage}{1.0\textwidth}
\begin{center}
\caption{SGR and AXP burst/afterglow energetics$^{a}$ and temporal decay indices.} 
\vspace{10pt}
\begin{tabular}{ccccc} \hline \hline

 & SGR~1900$+$14$^{b}$ & SGR~1627$-$41 & 1E~2259$+$586$^{c}$ & 1E~2259$+$586$^{d}$   \\\hline

Burst Energy (ergs; 25$-$100 keV) & 1 $\times$ 10$^{44}$ & 
	4 $\times$ 10$^{42}$ & 1 $\times$ 10$^{38}$ & 1 $\times$ 10$^{38}$  \\
Tail Energy (ergs; 2$-$10 keV)    & 2 $\times$ 10$^{42}$ & 
	3 $\times$ 10$^{42}$ & 3 $\times$ 10$^{39}$ & 2 $\times$ 10$^{41}$ \\
Decay Index   &  $-$0.71  &  $-$0.47  &  $> -6.6$  &  $-$0.22     \\

\hline\hline
\end{tabular}
\end{center}
\noindent$^{a}$ The following distances are assumed for conversions to energy:
15 kpc for SGR~1900$+$14 (Vrba et al.\ 2000); 11 kpc for SGR~1627$-$41 
(Corbel et al.\ 1999); and 3.0 kpc for 1E~2259$+$586 (Kothes et al.\ 2002). \\
\noindent$^{b}$ The values given here correspond only to those measured for the
1998 August 27 flare and its associated afterglow. \\
\noindent$^{c}$ The tail energy given here is derived from the excess 
persistent emission observed during the burst observations only.  Because of 
the ambiguities in the epoch determination, an accurate decay index cannot be
measured.  See \S3.6 for details. \\
\noindent$^{d}$ The tail energy given here is derived from the excess 
persistent emission observed during the long time scale decay.  See \S3.6 for 
details. \\
\end{minipage}\hfill
\end{table}

The burst and persistent emission energetics and temporal decay indices for all
three sources, 1E~2259$+$586, SGR~1900$+$14 and SGR~1627$-$41 are listed in
Table 5.  Clearly, there are large differences in both the decay index and the
ratio of burst to persistent emission energetics among the three sources.  One
possibility is that the response of the persistent X-ray flux to (or when
accompanied by) burst activity in SGRs and AXPs varies within the group. 
Alternatively (or in addition), there may be two components to the SGR flux
decay similar to the AXP flux evolution and the values listed in Table 6 for
the SGRs reflect a mixture of the two components which contributes to the
quantitative differences with the AXP values.  In fact, the flux from
SGR~1900$+$14 in the months following the end of the 40 day afterglow was
enhanced relative to the pre-outburst level (Woods et al.\ 2001).  This
enhancement could be due to the persistent, low-level burst activity observed
during this time interval or perhaps due to a slower flux decay component,
analagous to the slow flux decay seen in 1E~2259$+$586.  Unfortunately, there
are no reported spectral measurements of SGR~1900$+$14 during this epoch to
help distinguish between the two possibilities.  For SGR~1627$-$41, the initial
follow-up observation was not performed until $\sim$50 days after the primary
outburst (Woods et al.\ 1999a), therefore, any short-lived transient flux decay
would have been missed.  Thus, it appears that two flux components may possibly
be present in all SGR/AXP outbursts.  More rigorous follow-up spectral
measurements after future outbursts are needed to show whether or not this
behavior is ubiquitous.

\subsubsection{Comparison to SGR Outbursts: Pulse Properties}

SGR~1900$+$14 has shown one clear instance of a correlated change in the pulse
profile and pulsed fraction following the burst of 1998 August 29 (Lenters et
al.\ 2002).  Here, the pulsed fraction increased to a maximum value of
$\sim$20\% at $\sim$200 s following the burst, then rapidly decayed back to the
pre-burst value of $\sim$12\% within 10$^{4}$ s.  During the first $\sim$100 s,
the pulse shape showed large changes relative to the pre-burst profile, and
similar to the pulsed fraction recovery, recovered fully (within the errors)
during the next 10$^{4}$ s.  In 1E~2259$+$586, we observe comparable changes in
both the pulse shape and pulsed fraction at early times ($\lesssim$6 days)
during the outburst.  However, the time scale is significantly longer for
1E~2259$+$586 and the pulsed fraction decreased rather than increased.

The X-ray pulse profile of SGR~1900$+$14 changed from a complex, multi-peaked
shape before the giant flare of 1998 August 27 to a simple, nearly sinusoidal
profile following the flare (Woods et al.\ 2001).  This change has persisted
for years following the outburst and has yet to recover ({G\"o\u{g}\"u\c{s}} et
al.\ 2002).  The constraints on any change in pulsed fraction in SGR~1900$+$14
following this flare are weak, as the first reported pulsed fraction
measurement was made at 19 days following the flare, consistent with the
pre-flare value (Woods et al.\ 2001).  In 1E~2259$+$586, there was a
significant change in the pulse profile in which the power in the fundamental
frequency increased relative to the higher harmonics, analogous to what was
observed in SGR~1900$+$14.  The difference in pulse shape from before to
during/following the burst activity was not as profound as the change seen in
SGR~1900$+$14, however, the burst activity in 1E~2259$+$586 was not nearly as
energetic either.  Note also that the change in pulse profile of 1E~2259$+$586
was transient as the pulse shape at one year following the outburst is very
similar to the pre-outburst pulse profile.

\subsubsection{Physical Interpretation}

In spite of the quantitave differences mentioned above, many of the features of
the outburst in 1E~2259$+$586 are qualitatively very similar to those seen in
SGR outbursts -- specifically, the flux enhancement, spectral change, and
correlated change in pulse properties.  The similarities outlined above further
solidify the connection between SGRs and AXPs.  Combined with the similar burst
characteristics (Gavriil et al.\ 2003), this outburst in 1E~2259$+$586 shows
beyond any reasonable doubt that SGRs and AXPs are of the same nature, as
predicted uniquely by the magnetar model.

During the rapid initial flux decay, the blackbody radius was smaller ($\sim$1
km) than at all other times and the temperature was higher (0.8$-$1.7 keV). 
The thermal component of the spectrum suggests the existence of a hot spot that
is either cooling through its surface or is being heated by energetic particles
accelerated in the magnetosphere.  A localized hot spot will clearly result in
a change in the emitted radiation pattern (i.e.\ pulse profile) even if the
pulse shape is strongly modified by scattering in the magnetosphere (Thompson,
Lyutikov, \& Kulkarni 2002).  The reduction in pulsed fraction suggests that
the heated region is offset in angle from the locations on (or above) the
stellar surface that give rise to the two pulse maxima.

These observations do not allow us to distinguish convincingly between
magnetospheric emission, and passive cooling of an impulsively heated crust, as
the underlying source of the transient X-ray emission. The rapid flux decay
could have a very different time scaling from what is  observed following
intermediate energy bursts in SGR~1900$+$14, and any  change in the magnetic
field of 1E~2259$+$586 may have occured too  gradually to generate a bright
X-ray outburst.   A large current density  will be excited in the magnetosphere
above regions of strong crustal  shear, but the mechanism by which this current
is damped depends  on how rapidly it is excited (Thompson \& Duncan 2001).  The
rapid flux  decay as measured is consistent with a gradually diminishing creep 
within a small region of the crust over a period of $\sim 10^5$ s.

It is also worth considering whether this part of the X-ray transient is the
direct aftermath of a more energetic burst.  The short-lived afterglows
detected after SGR bursts of intermediate energy have a simple explanation as
the cooling of a pair-rich surface layer heated by a high-energy burst (Ibrahim
et al.\ 2001),  and as such are valuable probes of the burst mechanism.  As
estimated above, the minimum energy of such a burst is $\sim3.0\times 10^{41}$ 
($e_{\rm afterglow}/10^{-2})^{-1}$ ergs, where $e_{\rm afterglow}$ is afterglow
energy divided by the burst energy.  Konus data (S.~Golenetskii, private
communication) sets an upper bound of $\sim2\times 10^{38}$ ergs for bursts
with duration less than 3 s and $\sim2\times 10^{38}$ ($\Delta t$/3 s)$^{1/2}$
ergs for bursts with duration $3 < \Delta t < 1000$ s.  Therefore, a much
higher afterglow efficiency is required for 1E~2259$+$586 than is inferred for
the intermediate and giant SGR~1900$+$14 flares.

Is there any evidence for such extended energy release during SGR outbursts?
SGR~1900$+$14 has, in fact, been observed to radiate the same energy over time
scales ranging from $\sim 100$ s in the declining tail of the August 27 flare
(Feroci et al.\ 2001), down to $\sim 2$ sec in the August 29 burst that
followed it.  Thus, the putative burst from 1E~2259$+$586 would necessarily
have a much longer duration than any (of an equivalent energy) yet observed
from an SGR.  Although the  magnetospheric opacity scales as $B^{-2}$ and is
expected to be lower in 1E 2259$+$586 than in the actively bursting SGRs, the
luminosity of a passively cooling, trapped fireball would almost certainly
exceed the bound derived from Konus (e.g. for $B\sim B_{\rm QED}$).

We now comment on the relative importance of crustal heating and enhanced
magnetospheric emission for the slow decay component of the outburst of
1E~2259$+$586.  Bulk heating of the crust of a magnetar can power an excess
heat flux from its surface  for a year or longer, and has been proposed as the
explanation for the quasi-power-law flux decay seen in SGR~1900$+$14
(Lyubarsky, Eichler \& Thompson 2002) and SGR~1627$-$41 (Kouveliotou et al.\
2003).  In each case, an initial energy deposition of 10$^{44}$ ergs was
assumed.  For SGR~1900$+$14, there was in fact a giant flare preceding the flux
enhancement that emitted $\sim$10$^{44}$ ergs in soft gamma-rays.  For
SGR~1627$-$41, the burst energy output was more than one order of magnitude
less, so it was suggested that the process that generates the burst energy was
much less efficient in this source (Kouveliotou et al.\ 2003).   As discussed
previously, the relative energy released in burst and excess persistent
emission remains very uncertain in the case of 1E~2259$+$586.  Still, the
excess persistent energy output in the slow decay component (2 $\times$
10$^{41}$ ergs) was only a factor 10 lower than the output from the two SGRs,
yet there was not a {\it single} burst detected with an isotropic luminosity
above $\sim$1 $\times$ 10$^{39}$ ergs s$^{-1}$ during this outburst.  Thus, if
we ascribe the slow flux decay seen in these three sources to deep heating of
the crust of a magnetar, then it follows that 1E~2259$+$586 is almost certainly
less efficient in producing burst emission than either of the other two SGRs.
Note also that the similarity in the time scales for the relaxation of the
torque and the X-ray pulse profile must be coincidental if the slow decay
component of the X-ray emission is powered by crustal cooling.  

A small twist of the stellar crust, as is hypothesized to explain the glitch
characteristics (\S4.2.4), will impart a twist in the field lines that are
anchored to this patch of crust.  This drives a current along these twisted
field lines which ultimately produces X-ray emission whose luminosity is
proportional to the twist angle (Thompson et al.\ 2000; Thompson et al.\
2002).   However, following a twisting motion of the crust and external
magnetic field through an angle $\theta$, at most a fraction  $\sim 0.03
(\theta/0.01) (\theta_{\rm max}/10^{-3})^{-1} (B_{\rm dipole}/10^{14}~{\rm
G})^2$ of the energy released is stored in the external non-potential magnetic
field.  (Here $\theta_{\rm max}$ is the maximum strain that the crust can
sustain before yielding.) The majority of the energy is dissipated through the
deformation of the crust itself, because the shear modulus at the base of the
crust corresponds to a much stronger magnetic field, $B = (4\pi\mu)^{1/2}
\simeq 6\times 10^{15}$ G.   The amplitude of the required global twist (about
$\sim 10^{-2}$ radians) will only slightly change the optical depth to resonant
cyclotron scattering, or the external torque through a flaring out of the
external dipole field (Thompson et al.\ 2002).  Hence, this cannot explain the
large shape changes seen in the pulse profile within the first week following
the glitch.  It is possible, however, that the more subtle pulse profile
changes seen more than one week post-glitch can be explained as a slight change
in optical depth due to a twisting of (likely extended) field lines.  The
torque is more sensitive to a current localized on the most extended field
lines, as is discussed in \S4.2.5.

\subsection{Rotational Evolution}

The timing data clearly show that a large spin-up glitch occurred in
1E~2259$+$586 coincident with or perhaps shortly before the onset of burst
activity.  A portion of the frequency jump ($\sim$20\%) recovered in a
quasi-exponential manner within the next $\sim$60 days.  The long-term timing
in the year following the glitch shows a significant reduction in the spin-down
rate of this pulsar.  Below, we draw comparisons between the characteristics of
the 1E~2259$+$586 glitch with those of other neutron stars and offer possible
explanations for the observed behavior in 1E~2259$+$586.

\subsubsection{Comparison with Radio Pulsars: Overall Energetics}

A conservative bound on the change in rotational energy is obtained by treating
the star as a rotating solid body.  For a star having moment of inertia $I =
10^{45}$ g cm$^2$, one infers a change in the rotational energy of $3\times
10^{39}$ ergs, some two orders of magnitude smaller than the integrated excess
X-ray emission. If the glitch involves an exchange of angular momentum between
a superfluid with moment of inertia $I_{\rm sf}$, and the crust of the star
with moment of inertia $I_{\rm c}$, then the release of rotational energy is
$\Delta E_g \sim (2\pi)^2I_{\rm c}\Delta\nu(\nu_{\rm  sf}-\nu_{\rm c})$,  where
$\nu_{\rm sf}-\nu_{\rm c}$ is the equilibrium lag between the rotation
frequency of the superfluid and crust.  Evaluating this lag using the observed
$\sim 15$-d post-glitch relaxation time, one finds a value for $\Delta E_g$
that is smaller by a factor $10^{-3}(I_{\rm sf}/.02 I)^{-1}$.  Thus the
rotation of the star is manifestly not the source of energy for the transient
X-ray  emission.

The relaxation of elastic strains in the stellar crust is another  possible
source of energy.  The magnitude $\psi$ of these strains  could easily exceed
the equilibrium rotational bulge of the star, $(R_{\rm equator}-R_{\rm
pole})/R_{\rm NS} = \Omega^2 R_{\rm NS}^3/GM_{\rm NS} = 10^{-8}\,(P/{\rm 7
s})^{-2}$.    However, the energy stored in the strained crust is $\sim 3\times
10^{39}(\psi/10^{-4})^2$ ergs. The magnitude of the strain must, by way of
comparison, be smaller than $\sim 5\times 10^{-5}$ in the precessing pulsar
PSR~B1828$-$11 (Cutler, Ushomirsky, and Link 2003), whose spin is an order of
magnitude faster than that of 1E 2259$+$586.  We conclude that the observed
X-ray transient could be powered by the release of elastic energy only if
departure from spherical shape in the crust were comparable to that of a
neutron star with a $\sim 15$ ms spin period.

The remaining possiblity is that the transient X-ray emission is powered by
magnetic field decay.  The energy carried by the (exterior) magnetic field of
1E 2259$+$586 is comfortably larger: $8\times 10^{44}\,(B_{\rm
pole}/10^{14}~{\rm G})^2\, (R_{\rm NS}/10~{\rm km})^3$ ergs in the simplest
case of a centered  dipole.  This is clearly a lower bound to the total
magnetic energy of the star, as it does not account for the internal (e.g. 
toroidal) field.  The surface field (and magnetic energy) would also be 
significantly stronger,  by a factor $\sim (\Delta R/2R_{\rm NS})^{-3}$,  if
the dipole moment were offset from the center of the star  (with a separation
$\Delta R$ between the magnetic poles).

\subsubsection{Comparison with the AXP 1RXS~J1708$-$4009}

It is interesting to compare the 1E~2259$+$586\ glitch with those reported for
a different AXP, 1RXS~1708$-$4009.  This AXP has shown two glitches, one in
1999 (Kaspi, Lackey \& Chakrabarty 2000) and one in 2001 (Kaspi \& Gavriil
2003; Dall'Osso et al. 2003).  The first glitch was similar to those seen in
the Vela and other radio pulsars; it showed a step $\Delta \nu / \nu = 6 \times
10^{-7}$ with $Q = 0$, and $\Delta \dot{\nu}/\dot{\nu} \simeq 1$\%, such that
the magnitude of the spin-down increased.  The 2001 glitch, however, was
markedly different, and in some ways resembled the glitch seen in
1E~2259$+$586.  Specifically, it showed $\Delta \nu / \nu = 1.4 \times
10^{-7}$, which completely recovered exponentially on a time scale of 50~days
(i.e.\ $Q \approx 1$).  Whether it suffered a step in $\dot{\nu}$ depends on
the interpretation of an apparent $\ddot{\nu}$ detected between glitches; this
is discussed in more detail by Kaspi \& Gavriil (2003) and Dall'Osso et al.
(2003).  As argued by Kaspi \& Gavriil (2003), this glitch may have indicated
that 1RXS~1708$-$4009 underwent bursting activity some time in the few week
interval between monitoring observations that straddled the glitch epoch, but
which went unobserved.  However, no flux or pulse profile changes comparable to
those seen for 1E~2259$+$586 still weeks after its outburst were seen for
1RXS~1708$-$4009.  Thus, any radiative changes would have had to decay on
shorter time scales than seen in 1E~2259$+$586, and in particular on shorter
time scales than the exponential glitch relaxation time scale.  This must be
kept in mind when considering the coincidence in time scales of the
1E~2259$+$586 rotational exponential decay and the average pulse profile's
relaxation to its pre-outburst morphology.

\subsubsection{Comparison to SGRs}

From pulse frequency measurements leading up to and following the August 27
flare, it was determined that SGR~1900$+$14 likely underwent rapid spin {\it
down} ($\Delta \nu/\nu \sim -1 \times 10^{-4}$) at or near the time of the
flare (Woods et al.\ 1999b).  An analysis of the pulsations during the flare
itself (Palmer 2002) confirmed the existence of a timing anomaly, but only an
upper limit on the time scale for the frequency change was measured
($\lesssim$1 day).  For 1E~2259$+$586, we detect a sudden increase in the
rotation rate, as opposed to a decrease, in addition to a much smaller
magnitude as that inferred for SGR~1900$+$14.  In general, the available limits
on glitches (of either sign) in SGRs are poor.  This is a result of fewer and
less extensive phase coherent timing solutions for these sources due to their
strong timing noise (Woods et al.\ 2002) and weaker pulsed signals relative to
most AXPs.  Given the observational limitations for monitoring glitch activity
in the SGRs, it is not clear whether the glitch that accompanied the outburst
in 1E~2259$+$586 is common to SGR outbursts.

Although there are not any lengthy phase coherent solutions bridging the
boundary between quiescent and burst active states in SGRs to allow for
identification of glitches, there are sufficient observations during and
following a few outbursts to search for transient changes in spin-down rate. 
During the first two months following the outburst of 1E~2259$+$586, the
frequency derivative may have changed sign (assuming the pulse shape changes
did not affect our phase measurements, see \S3.1) then rapidly switched back to
spin down with a higher magnitude compared to pre-outburst.  The average
spin-down rate during the first two months after the glitch in 1E~2259$+$586
was approximately double the pre-outburst value.  Similar behavior has not been
seen during or following SGR outbursts.  The fairly well sampled frequency
histories of two SGRs (1900$+$14 and 1806$-$20) show that there is no direct
correlation between burst activity and enhancements in the spin-down rate
during or shortly following the outburst (Woods et al.\ 2002).  In fact, the
spin-down rate of SGR~1900$+$14 in the $\sim$40 days following the 1998 August
27 flare was at its lowest (in magnitude) historical value (Woods et al.\
1999b, 2002).

\subsubsection{Post-Glitch Relaxation}

In radio pulsars, post-glitch relaxation is almost always observed, and can be
classified into two regimes:  one in which the initial frequency jump relaxes
exponentially relatively quickly (i.e.\ time scales of hours to weeks), and one
in which the frequency jump does not relax, and which is generally accompanied
by a permanent or very slowly (time scales of years) relaxing increase in the
magnitude of the frequency derivative.  Many radio pulsar glitches show both
such behaviors (e.g. Shemar \& Lyne 1996, Wang et al.\ 2000).  These are today
generally interpreted in terms of vortex creep theory, in which the glitches
are due to sudden unpinning of angular momentum vortices in the neutron star
crustal superfluid (Anderson \& Itoh 1975).  The latter, in the model, rotates
faster than the crust, but in equilibrium is loosely coupled to it via pinning
of vortices to crustal lattice sites.  In this picture, the vortices creep
outward slowly due to thermal activation (Alpar et al.\ 1984a,b), slowly
transferring angular momentum to the crust. In equilibrium, the crust, slowed
down by the external torque of magnetic braking, shares an identical spin-down
rate with the superfluid, in spite of a non-zero angular velocity differential
between them.  Vortex creep theory identifies the two observational post-glitch
relaxation regimes with different pinning regions (Alpar et al.\ 1989, 1993): 
the exponential relaxation is associated with superfluid regions with
temperatures high compared to pinning energies, so that equilibrium can be
established with only a small angular velocity lag.  This linear regime
relaxation is mathematically equivalent to the original two-component model for
the crust-superfluid interaction suggested by Baym et al.\ (1969) which did not
incorporate the more recent vortex pinning theory.  When the temperature is low
compared to the pinning energy, a large lag is needed to establish equilibrium;
this is the non-linear regime and may be responsible for the longer-timescale
relaxations of spin-down rate enhancements.

For 1E~2259$+$586, qualitatively, we detect both such classical glitch
behaviors as well.  However, quantitatively, the post-glitch relaxation of
1E~2259$+$586\ is very different from that seen in radio pulsars. Specifically,
the exponential decay time is longer than what is observed in most pulsar
glitches, and the $Q$ value, the fraction of the glitch that heals, is also
large by pulsar standards, though neither value is extreme.  What is extreme is
the combination:  the net effect is that the pulsar, at one month post-glitch,
spins down for a couple weeks with more than {\it double} its long-term
pre-outburst spin-down rate. In radio pulsars, the spin-down enhancement is
typically a few percent, nearly two orders of magnitude smaller.  This is
because the spin-down enhancement is due to a temporarily reduced moment of
inertia, as a linear component of the crustal neutron superfluid re-establishes
equilibrium, and this superfluid has at most a few percent of the moment of
inertia of the crust.  As discussed by Kaspi et al. (2003), a spin-down
enhancement of order unity would imply that  {\it half} of the stellar moment
of inertia was effectively decoupled from the crust following the glitch, much
more than ought to be present in the crustal superfluid.  Indeed the crust
itself has a relatively small moment of inertia compared to the core, which is
thought to be coupled to the crust on short time scales (Alpar, Langer \& Sauls
1984; Alpar \& Sauls 1988).   Hence, the post-glitch relaxation of
1E~2259$+$586\ could imply a core decoupling.  This would provide the best
evidence yet for the existence of core superfluid.

However, the observed post-glitch relaxation can be understood in terms of
conventional crustal superfluid if it is assumed that the superfluid/crust
angular velocity lag temporarily {\it reversed} at the time of the glitch
(as was noted by Alpar, Pines, \& Cheng 1990 in the case of the 1988 Christmas
glitch of the Vela pulsar).  The standard two-component theory has as
equations of motion for the crust (including
everything coupled to it on short time scales), and superfluid:
\begin{equation}
I_{\rm c} \dot{\Omega}_{\rm c} = \frac{I_{\rm sf} (\Omega_{\rm sf} - 
\Omega_{\rm c})}{\tau} - N_{\rm ex},
\end{equation}
and
\begin{equation}
I_{\rm sf} \dot{\Omega}_{\rm sf} = - \frac{I_{\rm sf} (\Omega_{\rm sf} - 
\Omega_{\rm c})}{\tau},
\end{equation}
where $N_{ex}$ is the external torque on the crust and $\tau$ is the 
crust-superfluid energy dissipation rate.  The dissipation rate
can be estimated from the post-glitch exponential recovery time scale.
From these equations, it follows that the change in the crustal angular frequency
derivative is given by
\begin{equation}
\Delta \dot{\Omega}_{\rm c} = \frac{I_{\rm sf}}{I_{\rm c}} 
\frac{\Delta (\Omega_{\rm sf} - \Omega_{\rm c})}{\tau},
\end{equation}
or
\begin{equation}
\Delta (\Omega_{\rm sf} - \Omega_{\rm c}) = \Delta \dot{\Omega}_{\rm c} 
\frac{I_{\rm c}}{I_{\rm sf}} \tau.
\end{equation}

For $\tau \simeq 15$ days, $I_{\rm sf}/I_{\rm c} \simeq 0.01$ and  $\Delta
\dot{\Omega}_{\rm c} \simeq 2\pi \times 10^{-14}$~rad~s$^{-2}$, we find $\Delta
(\Omega_{\rm sf} - \Omega_{\rm c}) \simeq 8 \times 10^{-6}$~rad~s$^{-1}$. Since
$\Delta (\Omega_{\rm sf} - \Omega_{\rm c}) = \Delta \Omega_{\rm sf} - \Delta
\Omega_{\rm c}$ and $\Delta \Omega_{\rm c} = 3 \times 10^{-6}$~rad~s$^{-1}$, we
have that $\Delta \Omega_{\rm sf} \simeq 5 \times 10^{-6}$~rad~s$^{-1}$.  As
the lag was initially small and $\Omega_{\rm c} \simeq 0.9$~rad~s$^{-1}$,  this
implies the superfluid angular velocity changed by only a part in 200,000. 
However, this change was about twice the observed change in the crust.  Thus,
at the time of the glitch, in this picture, the superfluid was temporarily spun
down twice as much as the crust spun up.  The observed enhanced spin-down
post-glitch is then due to the crust transferring angular momentum back {\it
to} the superfluid, as the latter attempts to arrange its vortices so as to
re-establish equilibrium.

The overall activity in glitches is a measure of the moment
of inertia of the internal flywheel which gives up angular momentum
during glitch events (Link, Epstein \& Lattimer 1999).  The 
activity parameter is obtained by summing over all glitches measured 
over a long time interval $T$.  If the frequency lag between the
superfluid and the star is sourced only by the external torque,
then one has
\begin{equation}
{1\over \langle\dot{\nu}\rangle T}\sum_i \Delta\nu_i \leq
{I_{\rm sf}\over I}
\end{equation}
In the case of 1E 2259+586, we estimate $\Delta \nu/\nu = 4\times 10^{-6}$ over
an interval of $T = 10$ yrs (the last outburst of 1E~2259$+$586 being estimated
to have occured 10 years earlier [Iwasawa et al. 1992]; see also \S4.3).  The
resulting activity parameter is $\sim 0.1$, several times larger than what is
measured in glitching radio pulsars (Link et al.\ 1999).  Either $I_{\rm sf}$
is proportionately larger, or the spin rate of the superfluid is being reduced
by some mechanism other than vortex line unpinning and thermal creep.  For
example, smooth deformations of the stellar crust, driven by magnetic stresses,
can have the latter effect (Thompson et al.\ 2000).

A variety of models have been proposed to explain the origin of  glitches in
radio pulsars.   In most models, the rotational lag between  the superfluid and
the crust is the principle source of free energy. The frequency lag may, as a
result, decrease in magnitude during the  glitch -- but it is more difficult to
reverse its sign. For example, Ruderman (1991) suggested that the tension of
the superfluid vortices would fracture the crust, thereby allowing an outward
shift in the positions of the vortices.  Link and  Epstein (1996) noted that
any deposition of heat (of the order of $\sim 10^{41}-10^{42}$ ergs in the deep
crust) would accelerate the creep rate of the pinned vortex lines, thereby
causing a transient spin down of the superfluid, and a spin up of the rest of
the star.  Our observation of enhanced X-ray emission from 1E~2259$+$586 is,
indeed, consistent with this level of heat deposition  in the crust.  But it
appears that this cannot be the entire mechanism responsible for the observed
glitch, because the vortex creep rate goes to  zero as the spin rate of the
superfluid approaches that of the  crustal lattice.

In a slowly rotating neutron star, magnetic stresses can act on time scales
much shorter than the external torque.  Glitch-like events will result either
from sudden fractures of the crust (Thompson \& Duncan 1993); or from more
gradual plastic deformations during which the vortices remain pinned (Thompson
et al.\ 2000).    The lowest energy deformations of this type are torsional.
Consider, for example, a twisting motion of a circular patch of the crust that
is offset in azimuth from the rotation  axis.  Starting from a uniform
distribution, more vortices are  advected outward than inward (due to the
curvature of the stellar  surface).  A twist of the order of $10^{-2}$ radians
is enough to  provide the required spin down of the crustal superfluid in
1E~2259$+$586  (eq. [12] of Thompson et al.\ 2000).  By forcing the superfluid
to rotate more slowly than it otherwise would, this mechanism also provides an
explanation for the unusually large glitch activity that we inferred for
1E~2259$+$586 (see also Heyl \& Hernquist 1999).

Finally, as noted by Kaspi et al. (2003), the spin-down rate enhancement could
also be due to a magnetospheric restructuring which causes the external torque
to approximately double.  However, given that the spin up itself is inescapably
enabled by an internal transfer of angular momentum, it would have to be a
suprising coincidence that the external torque change offset, to within a
factor of 4-5, the internally generated frequency jump.  This of course does
not preclude some magnetospheric restructuring; indeed the residual pulse
profile change following the initial rapid flux decay and the post-outburst
infrared enhancement (Kaspi et al. 2003) suggest that likely occured.  However,
it seems unlikely that a magnetospheric restructuring is the cause of the
enhanced spin down just post-outburst.

\subsubsection{Post-Glitch Long-Term Rotation}

Our glitch fitting clearly measured a long-term post-glitch change in the
spin-down rate.  Interestingly, unlike that seen in a number of radio pulsars
to date, the magnitude of the spin-down torque {\it decreased}.   (In the case
of the Crab pulsar, for example, the cumulative effect of the torque changes
observed following some glitches  is to increase the rate of spin-down by
0.07\% over 23~yr of observations [Lyne, et al.\ 1993]). If due to a superfluid
effect, the conventionally  observed long-term increase in magnitude of
$\dot{\Omega_{\rm c}}$ represents a  reduction in the moment of inertia of the
part of the neutron star that  is coupled to its crust.  This suggests that 
some fraction of the crustal superfluid is tightly pinned not creeping outward,
hence decoupled from the stellar spin-down (Alpar et al.\ 1994), with this
fraction increasing at the glitch. The opposite must therefore be true of
1E~2259$+$586 in this scenario.  It has alternatively been suggested that the
effect is due to a secular change in the magnetic moment (Ruderman 1991); or to
a change in the electromagnetic torque acting on the star, driven by a
reorientation of the magnetic moment with respect to the angular velocity of
the star (Link \& Epstein 1997).

We now comment on the relative merits of a change in internal vs.\ external 
torque in the case of 1E~2259$+$586.  The sign of the torque change is
interesting, because there is independent evidence that the spin-down torque of
1E~2259$+$586 has decayed over time: its characteristic age, $\tau_c = 2 \times
10^5$ yrs, is an order of magnitude greater than that of the SNR CTB~109
(estimated radius 12 pc; Kothes et al.\ 2002).  A change in vortex creep rate
with  a time scale $\tau$ much longer than 1 yr, but much less than the
characteristic age $\tau_c = 2\times 10^5$ yrs, would appear as a permanent
change in braking torque. The required moment of inertia of the superfluid
component would be $I_{\rm sf}/I = 0.05(\tau/\tau_c)$ -- comfortably below the
fraction inferred for glitching radio pulsars (Link, Epstein \& Lattimer 1999)
unless the creep time scale were more than a million times longer than the
observed 15 day relaxation.  The simplest explanation for the required change
is the external torque -- a $\sim 2.5$\% reduction in the magnetic moment --
can be discounted.  If, for example, the external magnetic field of
1E~2259$+$586 were a centered dipole, the total field energy outside the star
would amount to some $10^{45}$ ergs, and the reduction in field energy would be
a few hundred times  larger that the excess X-ray emission we have observed.  

Nonetheless, changes in external torque can result from more subtle effects. If
the magnetosphere is globally twisted due to the action of internal magnetic
stresses, the resulting expansion of the poloidal field lines increases the
current crossing the speed of light cylinder, and therefore yields a greater
torque than would exist from a simple dipolar field rotating in a vacuum
(Thompson et al.\ 2002).  Much of this effect results from the closed field
lines which extend far from the star to a large radius $R_{\rm max}$, are
anchored in a small fraction $\sim {1\over 2}(R_{\rm max}/R_{\rm NS})^{-1}$ of
its surface area, and which contain only a small fraction $\sim (R_{\rm
max}/R_{\rm NS})^{-3}$ of the energy in the non-potential (toroidal) field
outside the star. A slight relaxation in the twist of the closed field lines is
therefore  consistent with the observed energetic output following the glitch,
if that reconfiguration occurs close to the magnetic poles.   Persistent
seismic activity in a magnetar will also modulate the spin-down torque, through
a coupling of crustal shear waves to the magnetosphere which in turns drives a
particle wind (Thompson \&  Blaes 1998; Harding, Contopoulos \& Kazanas 1999;
Thompson et al.\ 2000).   In this case, the observed change in torque
corresponds to a very small  reduction in the particle luminosity, by only
$\sim 0.1 I\Omega \dot\Omega \sim 10^{31}$ erg s$^{-1}$. 

Finally, a change in torque could result from the excitation of precession
during the glitch, followed by a permanent reorientation of the magnetic axis
relative to the rotation axis of the star (Link \&  Epstein 1997).  That would
require a precession angle of at least a few degrees, which is uncomfortably
large if the non-spherical shape of the star is predominantly due to its
magnetic field.  The fractional difference in the sizes of the principal
moments is $\sim 10^{-6}\,[\langle B^2\rangle/(10^{15}~{\rm G})^2]$ (Wasserman
2003).  However, the observed energy of the X-ray transient  corresponds to
less than $\sim 10^{-3}$ of the total magnetic energy of  the star (\S 4.2.1) 
and, thus, to a proportionately small change in the  shape of the star.

\subsection{Implications for the Number of Active Magnetars}

As suggested by Kaspi et al.\ (2003), past reports of flux variability (Iwasawa
et al.\ 1992; Baykal \& Swank 1996), pulse profile changes (Iwasawa et al.\
1992), and glitch activity (Heyl \& Hernquist 1999) in 1E~2259$+$586 likely
indicate previous episodes of burst activity in this source.  The best example
of an inferred outburst from 1E~2259$+$586 comes from a pair of {\it Ginga}
observations in 1989 and 1990.  Between 1989 and 1990, the pulse profile of
1E~2259$+$586 changed drastically, the flux increased by more than a factor 2,
and the measured frequency showed evidence of a glitch.  The close similarities
with the well studied outburst of 2002 June strongly suggests that a bursting
episode from 1E~2259$+$586 preceded the 1990 {\it Ginga} observation by roughly
one week.  Similar reports of flux variability and timing anomalies in the AXP
1E~1048.1$-$5937 (e.g.\ Oosterbroek et al.\ 1998) suggests past episodes of
burst activity in this source as well.

What is most intruiging about the detected and inferred outbursts in AXPs is
that none has been detected with large FOV gamma-ray detectors (e.g.\ BATSE,
Konus, Ulysses, etc.) that traditionally detect burst active episodes in SGRs. 
This shows that we are missing low intensity SGR-like outbursts from magnetar
candidates in our Galaxy (i.e.\ the SGRs are a sensitivity limited sample).

The number of active SGRs in our Galaxy was calculated previously to be
$\sim$10 (Kouveliotou et al.\ 1993).  A key assumption that was made in this
estimation was that SGR outbursts are easily detected (i.e.\ there are always
bright bursts within a given outburst that would trigger at least some of the
large FOV gamma-ray instruments).  The discovery of low-intensity SGR-like
outbursts in AXPs clearly invalidates that assumption, making the estimate of
10 a lower limit only.

The future prospects for detecting weak SGR-like outbursts and addressing the
question of the number of active SGRs in our Galaxy is promising.  Up until
2000 June, BATSE was the most sensitive large FOV detector for SGR burst
emissions.  The impending {\it Swift} mission and its sensitive BAT detector
will be $\sim$20 times more sensitive than BATSE was (D.~Band, private
communication).  Just as the {\it RXTE} PCA opened our eyes to the larger
population of low-luminosity SGR bursts (e.g.\ {G\"o\u{g}\"u\c{s}} et al.\
1999), perhaps {\it Swift} will reveal a previously unknown source population
of dimmer SGRs.

\section{Summary}

In 2002 June, the AXP 1E~2259$+$586 was observed to emit more than 80 SGR-like
bursts.  Accompanying this outburst were several changes in the persistent
X-ray emission properties of the source, many of which are similar to what has
been seen in SGRs, thus further blurring the distinction between the two
classes.  We have quantified the observed changes using data obtained with {\it
XMM-Newton} and {\it RXTE}.  In particular, we found the following:

\begin{itemize}

\item{The flux increased by more than an order of magnitude and showed two
decay components.  The first component decayed rapidly within the first day of
the outburst while the second decayed much more slowly as a power law in time
according to $t^{-0.22}$.}

\item{The X-ray spectrum hardened during the outburst, but almost fully
recovered within three days.  The spectrum at 21 days past the burst activity
was significantly harder than at 7 days pre-outburst, but fully consistent with
historical spectral measurements of this source in quiescence.}

\item{The phase dependence of the energy spectrum changed from before to after
the outburst.  One week prior to the outburst, we observe significant
variablility in the photon index, but not the blackbody temperature.  Three
days following the outburst, the dependence of the photon index on pulse phase
flattened significantly.}

\item{The pulse profile changed suddenly during the observation containing the
burst activity where much of the power moved to the fundamental frequency.  The
pulse profile rapidly returned to near its pre-outburst shape within one week,
and showed only very slow recovery thereafter.  As with the flux and spectral
changes, the recovery was not complete even after one year.}

\item{The pulsed fraction decreased during the outburst to $\sim$15\%, but
quickly recovered to the pre-outburst value of $\sim$23\% within six days.}

\item{1E~2259$+$586 suffered a glitch having an ordinary amplitude
($\Delta\nu_{\rm max}/\nu = (4.24 \pm  0.11) \times 10^{-6}$), but a unique
recovery.  Approximately 19\% of the glitch recovered on a time scale of
$\sim$16 days, although the recovery was not exactly exponential in form.
During the recovery, the torque was enhanced relative to the pre-outburst value
by a factor $\sim$2.  We detect a long-term $\sim$2\% {\it reduction} in the
spin-down rate following the glitch ($\Delta \dot{\nu} = 2.18 \pm 0.25 \times
10^{-16}$ Hz s$^{-1}$).}

\item{The measured glitch epoch preceedes the observed burst activity by
$\sim$12 hours.  Given the rapid flux decay during the outburst, the true onset
of this burst activity may have followed the glitch.}

\end{itemize}

\noindent The cumulative properties of the outburst in 1E~2259$+$586 lead us to
conclude that the star suffered some major event that was extended in time and
had two components, one tightly localized on the surface of the star (i.e.\ a
fracture or a series of fractures) and the second more broadly distributed
(possibly involving a smoother plastic change).  This event affected both the
superfluid interior and the magnetosphere.  The glitch points toward a
disturbance within the superfluid interior while the extended flux enhancement
and pulse profile change suggests an excitation of magnetospheric currents and
crustal heating.  Finally, we show that the lack of detection of AXP outbursts
with all-sky gamma-ray detectors suggests there exists a larger population of
active SGRs in our Galaxy than previously thought.

\acknowledgments{\noindent {\it Acknowledgements} --  }  We thank P.~Plucinsky
(PI of the CTB~109 observations) for providing {\it XMM-Newton} data on
1E~2259$+$586.  We would like to thank W.~Becker, R.~Epstein, M.~Finger,
A.~Harding, C.~Kouveliotou, B.~Link, F.~\"Ozel, D.~Psaltis, and I.~Wasserman
for useful discussions.  We thank J.~Swank and the {\it RXTE} scheduling team. 
We thank F.~Jansen for scheduling a ToO observation with {\it XMM-Newton}.  We
thank S.~Golenetskii and E.~Mazets for providing limits on burst emission from
1E~2259$+$586 using their Konus data set.  PMW acknowledges support through
NASA.  VMK is a Canada Research Chair, an NSERC Steacie Fellow, and a Fellow of
the CIAR.  Funding to VMK and FG was provided by an NSERC Discovery Grant and
Steacie Supplement, NATEQ, CIAR, and NASA.  CT is supported by the NSERC of
Canada.  JSH is supported under a Chandra Postdoctoral Fellowship Award.

\end{document}